\documentclass[%
reprint,
 amsmath,amssymb,
 aps,
]{revtex4-2}
\usepackage{graphicx}
\usepackage{subcaption}
\usepackage{caption}
\usepackage{subcaption}
\usepackage{lipsum}
\usepackage{gensymb}
\usepackage{dcolumn}
\usepackage{bm}

\begin{document}

\preprint{APS/123-QED}

\title{Engineering Phonons in Compositionally Complex Carbide Ceramics}
\author{Linu Malakkal$^1$}
\email{linu.malakkal@inl.gov}
\author{Jarin C French$^1$}%
\author{Lanh Trinh$^2$}%
\author{Kaustubh K Bawane$^1$}%
\author{Shuxiang Zhou$^1$}%
\author{Zilong Hua$^1$}%
\author{Lingfeng He$^3$}%
\author{Yongfeng Lu$^2$}%
\author{Bai Cui$^2$}%

\affiliation{
 $^1$Idaho National Laboratory, Idaho Falls, Idaho 83415, USA}

\affiliation{
 $^3$North Carolina State University, Rayleigh}
 
 \affiliation{
 $^2$University of Nebraska-Lincoln, Lincoln, NE 68588, USA}
 
\date{\today}

\begin{abstract}
In the pursuit of advanced ceramic materials with exceptional irradiation-resistance and high-temperature tolerance for nuclear applications, compositionally complex carbides (CCCs) have emerged as a highly promising class of candidate materials for extreme environments. In such conditions, critical material properties such as thermal stability, elasticity, thermal conductivity and thermodynamics behavior are predominantly influenced by phonons. In CCCs, pronounced cation disorder can lead to significant phonon scattering due to inherent mass and force constant variations, impacting these critical properties. In this study, we used \textit{ab initio} calculations to predict the phonon band structures and systematically explore the influence of mass and force constant variance on the phonon spectral function of CCCs with a rock salt structure, ranging from binary to five-metal component carbides. Our findings reveal that the selection and concentration of constituent elements can be strategically utilized to tune the phonon band structure, phonon bandgap and phonon scattering in CCCs, thereby enabling control over phonon-related properties. Additionally, we measured the thermal conductivity of some of these CCCs using the spatial-domain thermoreflectance technique.  Interestingly, the measured thermal conductivity of some of these CCCs indicates that five-component ceramics exhibit higher thermal conductivity than certain ternary and binary alloys. This observation contrasts with the expectation that greater cation disorder would result in more scattering and lower thermal conductivity. This intriguing result opens up the possibility of discovering CCCs with better thermal conductivity, presenting new opportunities for their application in extreme environments.
\end{abstract}

\maketitle
\section{\label{sec:level1}Introduction\protect}
The development of Generation-IV fission and future fusion reactors necessitates the identification of new materials capable of enduring extreme conditions such as high temperatures, high stress, significant neutron damage, and high transmutation gas sustainability \cite{Zinkle2013,Hosemann2018}. These materials must exhibit exceptional resistance to neutron irradiation damage, corrosive coolants, irradiation creep, and helium embrittlement, while maintaining superior mechanical and thermal properties. Traditional ceramic materials such as silicon carbide (SiC) \cite{SiC_review_Qiu_2020} and zirconium carbide (ZrC) \cite{ZrC_KATOH_2013}, have been favored for nuclear applications due to their superior tolerance to high-temperature environments compared to metallic alloys like stainless steels and nickel alloys. However, ZrC and SiC still face limitations in their resistance to irradiation damage and their thermal and mechanical performance \cite{Ulmer2015,Katoh2012}. Therefore, identifying new classes of materials that can perform under these extreme conditions is crucial.

Recently, the concept of multi-principal element has been used to design and synthesize novel ceramic materials known as compositionally complex carbides (CCCs) \cite{Zhou2018,Yan2018} that exhibit both irradiation resistance and high-temperature tolerance \cite{WANG_2020_irradiation}. CCCs, comprising combinations of zirconium carbide (ZrC), titanium carbide (TiC), tantalum carbide (TaC), hafnium carbide (HfC), tungsten carbide (WC), vanadium carbide (VC), and niobium carbide (NbC), have shown exceptional properties. These properties include outstanding thermal stability \cite{HARRINGTON_2019}, robust mechanical properties at both elevated and ambient temperatures \cite{DEMIRSKYI_2020_SPS, Wang_2020}, and enhanced resistance to corrosion and irradiation compared to their individual monocarbide counterparts \cite{MALINOVSKIS_2018, Wang_2020}. The impressive physical and chemical characteristics, along with their resistance to irradiation, can be attributed to the inherent compositional complexity, atomic disorder, and lattice distortion, as highlighted in a recent review \cite{Trinh2024_review}. These factors also play a crucial role in influencing fundamental processes such as defect formation, deformation behavior, and lattice thermal conductivity \cite{Trinh2024_review}. Designing materials for extreme environments necessitates accurate knowledge of their thermodynamic properties, phase stability, and thermal conductivity, all of which are related to phonons. Although there has been significant interest in the phonon band structure of transition metal carbides (TMCs), specifically ZrC, TiC, TaC, NbC, and HfC,  since the 1970s due to their remarkable thermal properties and potential for superconductivity \cite{smith1980_phonons_ZrC,TiC_phonon_PRB_2009,Pintschovius_1978_TiC_exp,Smith_PRL_1972,Smith_1970_TaC_HfC}, a fundamental understanding of the phonon band structure in these CCCs is still lacking. Unlike perfect crystals or ordered alloys, multi-component random alloys can induce significant phonon broadening and scattering due to inherent variances in force constants and mass. The vast compositional space available for CCCs allows for tuning phonon band structure thereby systematically controlling the thermal properties. However, experimentally determining the phonon band structure of the CCCs is challenging, necessitating a quantitative understanding of phonon properties through first-principles approaches. Therefore, in this work, we use first-principles calculations to determine the phonon band structures of CCCs (containing Zr, Ti, Ta, Nb, and Hf), aiming to provide technical guidance for designing and discovering new ceramic materials for extreme environments.

To model the phonon band structures of these complex materials, commonly adopted methods include the virtual crystal approximation (VCA) \cite{VCA_David_2000} and the coherent potential approximation (CPA) \cite{Jaros_1985}. However, these methods have limitations as they do not explicitly include the local chemical environment around each atom, which is critical for the quantitative evaluations of the physical properties of disordered systems \cite{Zunger_1991}. Therefore, in this work, we use the supercell method, which takes into account the local environment around each atom, including the local relaxation of atomic positions. When a disordered system is modeled by the supercell method, it generally lacks the crystallographic symmetry of the underlying structures. This makes it challenging to compare the phonon band structure calculated from the supercell model with experimental data, which typically describe the disordered system as having the crystallographic symmetry of its underlying structure. To represent the phonon dispersion spectra obtained by the supercell calculations as an effective band structure that can be compared to experiments, the band-unfolding method is employed \cite{Boykin_2007, Ku_PRL, Allen_2013, Yuji_mode_2017}. Among these band-unfolding methods, we used the method developed by Ikeda et al. \cite{Yuji_mode_2017} to study phonon properties in CCCs. This methodology unfolds the phonon band structures according to the contribution of the chemical elements, enabling us to analyze the role of each element in the CCCs phonon band structure.

Utilizing this methodology, we predicted the phonon dispersion spectra of the CCCs, and elucidated the role of each element by systematically altering the composition of the CCCs, starting with binaries, progressing through ternaries, quaternaries, and finally quinaries. The ceramics studied include the binary ceramics Zr$_{0.5}$Ti$_{0.5}$C, Zr$_{0.5}$Ta$_{0.5}$C, Zr$_{0.5}$Nb$_{0.5}$C, Ta$_{0.5}$Nb$_{0.5}$C, and Zr$_{0.5}$Hf$_{0.5}$C; the ternary ceramics Zr$_{0.33}$Ti$_{0.33}$Ta$_{0.33}$C, Zr$_{0.33}$Ti$_{0.33}$Hf$_{0.33}$C, Zr$_{0.33}$Ti$_{0.33}$Nb$_{0.33}$C and Zr$_{0.33}$Hf$_{0.33}$Nb$_{0.33}$C; the quaternary ceramics Zr$_{0.25}$Ti$_{0.25}$Ta$_{0.25}$Nb$_{0.25}$C and Zr$_{0.25}$Ti$_{0.25}$Ta$_{0.25}$Hf$_{0.25}$C; and the quinary ceramics Zr$_{0.2}$Ti$_{0.2}$Ta$_{0.2}$Nb$_{0.2}$Hf$_{0.2}$C. The contributions from each element to the phonon band structure and the broadening were calculated for all ceramics. We also provide the details of how each element modifies the phonon dispersion spectra of the CCCs. This systematic understanding of phonon band structure and broadening in CCCs will enable researchers to select elements in CCCs more effectively, offering the possibility to tune phonon broadening by adjusting the chemical complexity through varying the relative fractions of constituent elements. Beyond our primary goal of predicting the phonon band structure of CCCs, we present experimental measurements of the thermal conductivity of the quinary ceramic Zr$_{0.2}$Ti$_{0.2}$Ta$_{0.2}$Nb$_{0.2}$Hf$_{0.2}$C, as well as binary (Zr$_{0.5}$Hf$_{0.5}$C) and ternary Zr$_{0.33}$Hf$_{0.33}$Nb$_{0.33}$C ceramics at low temperatures using the spatial-domain thermoreflectance (SDTR) technique. Contrary to the expectation that greater cation disorder would result in lower thermal conductivity, our findings reveal that the quinary ceramic exhibits higher thermal conductivity than the binary and ternary ceramics. This finding opens new avenues for identifying CCCs with enhanced thermal properties, even among significant cation disorder.

\section{Computational and Experimental Details}
In previous studies, the synthesized CCCs were all MC-type carbides (M = transition metal, such as Ti, Zr, Hf, V, Nb, Ta, Mo, W), which exhibit a rock salt B1 structure with a space group of $Fm\bar{3}m$ \cite{Yan2018,Castle_SPS,Trinh2024_review}. Therefore, in simulations, the crystal structure of the CCCs was modeled as rock salt structure, with two atoms per unit cell, resulting in six phonon modes at any given q-point. All density functional theory (DFT) \cite{Kohn_1965} calculations in this work were performed using the projector-augmented-wave (PAW) \cite{blochl_1994} method implemented in the plane wave software package VASP \cite{Kresse_1996}, with the local density approximation \cite{Perdew_1981} (LDA) pseudopotential for the exchange and correlation. A plane wave cutoff energy of 550 eV was chosen for all calculations. Chemical disorder was simulated using the special quasi random structure (SQS) \cite{Zunger-sqs}. The SQS cells for the 2-,3-,4-,and 5-component ceramics contained 216 atoms each. The atomic coordinates of each supercell were optimized with energy convergence critera set at 10$^{-7}$ eV per atom while maintaining the cubic cell shape. Methfessel-Paxton smearing \cite{MP_smearing} with a smearing value of 0.1 eV was used. For the total energy convergence, a K-point grid of 2$\times$2$\times$2 was employed. The second-order force constants of the supercell models were calculated using the finite displacement method as implemented in the PHONOPY package \cite{TOGO2015}. The first principles derived force constant have been employed in the band unfolding method \cite{Boykin_2007, Ku_PRL, Allen_2013, Yuji_mode_2017}. In this method, phonon spectra of random ceramics are obtained by decomposing the phonon modes of SQS models on to the Brillouin zone of the underlying crystal structure according to their translational symmetry. Pair projected contributions of the spectral functions are derived by the recently developed mode decomposition technique and by the proposed modified projection scheme \cite{Yuji_mode_2017}. Finally, the phonon unfolding was performed using the upho code \cite{Yuji_mode_2017} in combination with PHONOPY \cite{TOGO2015}.

Commercial monocarbide powders of ZrC (99.5\%, $<$45 $\mu$m particle size), HfC (99.5\%, $<$45 $\mu$m), NbC (99.0\%, $<$45 $\mu$m), TaC (99.5\%, $<$45 $\mu$m), and TiC (99.5\%, $<$45 $\mu$m) purchased from Alfa Aesar were used as starting materials for synthesizing a quinary (Zr$_{0.2}$Ti$_{0.2}$Ta$_{0.2}$Hf$_{0.2}$Nb$_{0.2}$)C, binary (Zr$_{0.5}$Hf$_{0.5}$)C and ternary (Zr$_{0.33}$Hf$_{0.33}$Nb$_{0.33}$)C CCC ceramic. For each material, the constitute monocarbides were mixed in equimolar ratios and ball milled in an Argon atmosphere using 10-millimeter (mm) diameter stainless steel grinding balls (ball-to-powder mass ratio of 5:1) at 250 rotations per minute (rpm) for 8 hours in a high-energy planetary ball mill (Model Pulverisette 7, Fritsch GmbH). The ball milling process was stopped for 5 minutes every 60 minutes for cooling to prevent overheating. The sintering process was performed with a spark plasma sintering (SPS, Model 10-4, Thermal Technologies) at 2000$\degree$C for 10 minutes at a pressure of 30 MPa. The as-sintered samples were 10 mm in diameter and 3 mm thickness. The theoretical density of each material was calculated from the mass of the atoms in a unit cell and the lattice parameters of rock salt crystal structure determined by X-ray diffraction. The real density was measured  using Archimedes’ methods in water on a balance (AT201, Mettler Toledo), which was used to calculate the relative density of each sample: (Zr$_{0.5}$Hf$_{0.5}$)C (95.6$\%$), (Zr$_{0.33}$Hf$_{0.33}$Nb$_{0.33}$)C (96.3$\%$), and (Zr$_{0.2}$Ti$_{0.2}$Ta$_{0.2}$Hf$_{0.2}$Nb$_{0.2}$)C (88.9$\%$).

 The thermal conductivity of the selected CCCs were measured using the SDTR technique. SDTR is a pump-probe technique utilizing continuous-wave lasers, with the pump being amplitude-modulated. One advantage of this measurement method is its insensitivity to optical spot size \cite{Feser_2012}, which improves measurement reliability and reproducibility. More detailed information about this measurement technique and the 3D thermal wave model used to extract the thermal conductivity and thermal diffusivity from the data can be found in \cite{Hurley_2015,Hua_2012,Maznev_1995}. For our SDTR setup, we used 660-nm (Coherent OBIS 660 nm) and 532-nm (Coherent Verdi 532 nm) continuous-wave lasers for the pump and probe, respectively. The power arriving at the sample surface was measured as $\sim$3 mW for the pump and $\sim$0.3 mW for the probe. The periodic temperature variation in both time and spatial domain, introduced by the pump laser, is captured by the probe laser via thermoreflectance effect, and used for thermal property extraction. The laser beams are focused using a 50× long-working-distance objective lens, resulting in a $\sim$1 $\mu$m spot size. In order to create the necessary boundary condition to independently measure thermal conductivity and diffusivity and to improve the thermoreflectance effect, a thin layer of gold film with the thickness of 85 nm was coated on all samples, of which the thermal conductivity (at room temperature) was measured from a simultaneously coated reference BK7 sample. The thermal conductivity measurement was conducted at cryogenic temperatures ranging from 77 K to 300 K, with increments of 25 K (between 100 K-275 K). The sample was mounted inside a liquid nitrogen-cooled optical cryostat (Cryo Industries model XEM), with the temperature maintained at the intended temperature, exhibiting a maximum fluctuation of 0.3 K. The pressure within the cryostat chamber was kept below 5 mTorr, and the chamber was initially purged with ultrahigh-purity nitrogen to prevent ice condensation on the sample surface. At each temperature, a minimum of four sets of measurements were conducted with three different pump modulation frequencies: 20, 50, and 100 kHz. For each frequency measurement, the pump was scanned across the probe beam over a 10 $\mu$m distance in each direction. When fitting the thermal conductivity of the carbide samples at low temperature, both thermal conductivity and heat capacity of the gold thin film were corrected using literature values \cite{takahashi1986heat} and our previous measurements. Multiple data sets were collected to reduce the statistical uncertainty of the thermal conductivity measurements.
 
\section{Results and Discussion}

\begin{figure*}[!htb]
    \centering
    \begin{subfigure}[b]{0.3\textwidth}
        \includegraphics[width=\textwidth]{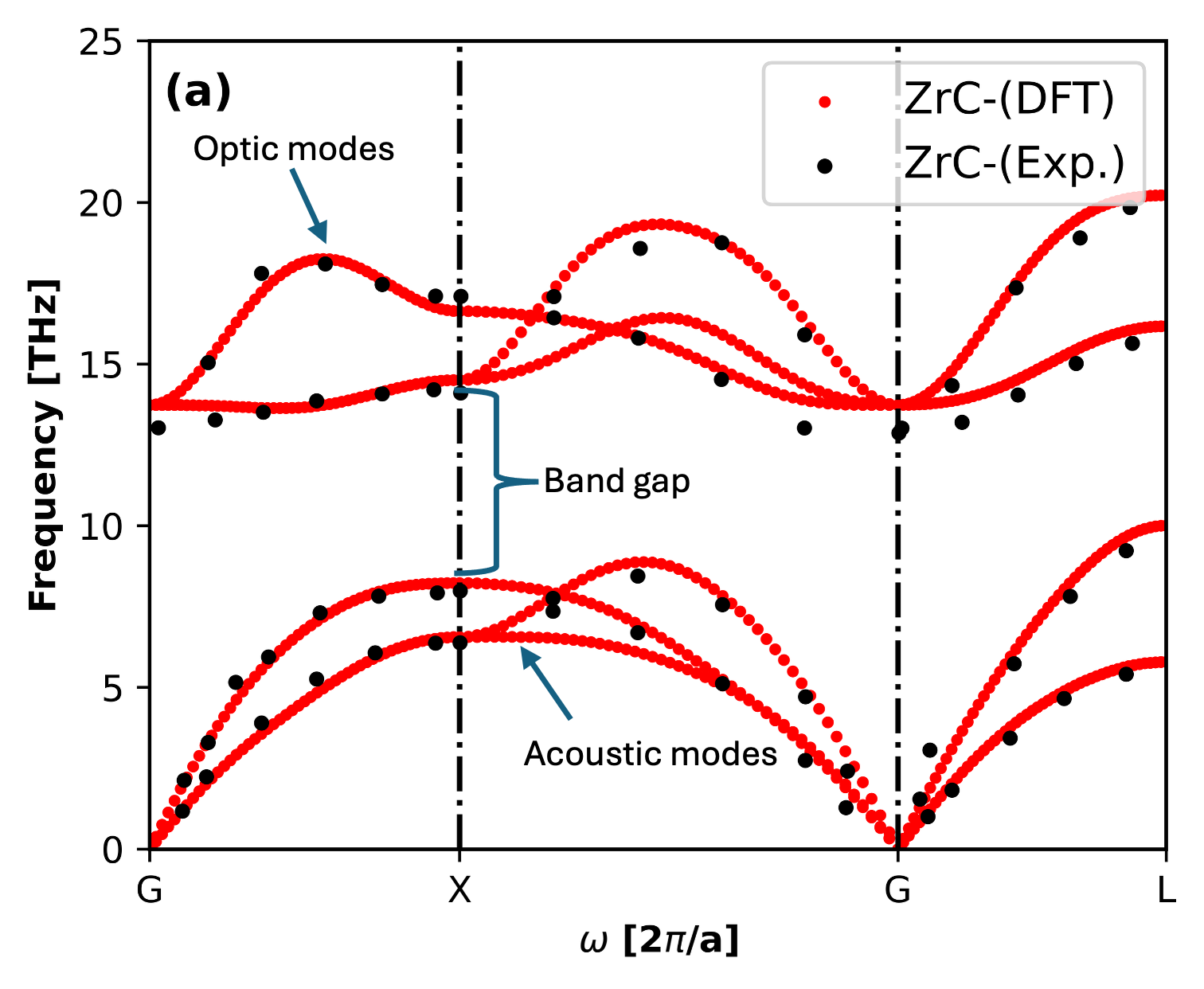}
    \end{subfigure}
    \begin{subfigure}[b]{0.3\textwidth}
        \includegraphics[width=\textwidth]{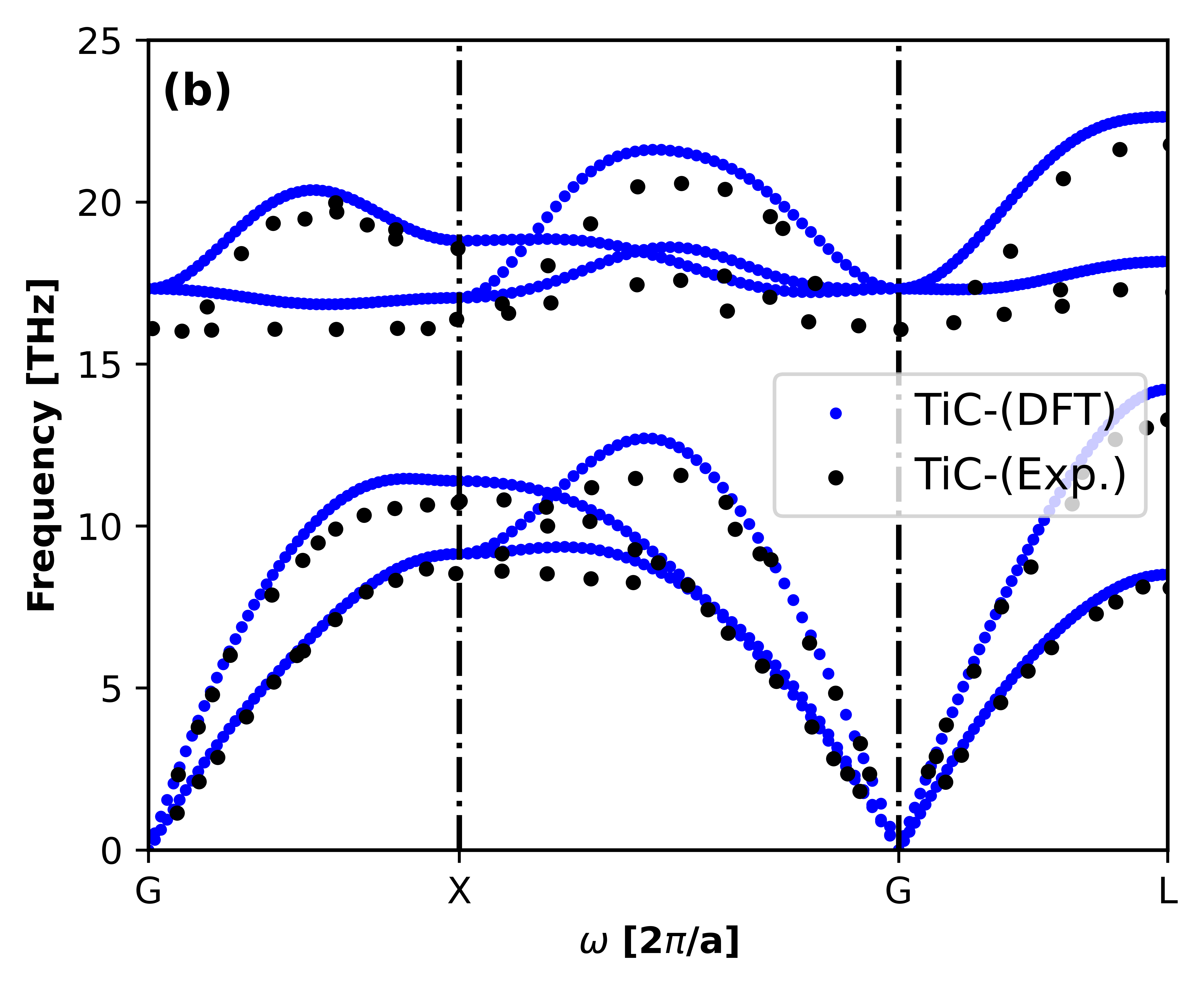}
    \end{subfigure}
    \begin{subfigure}[b]{0.3\textwidth}
        \includegraphics[width=\textwidth]{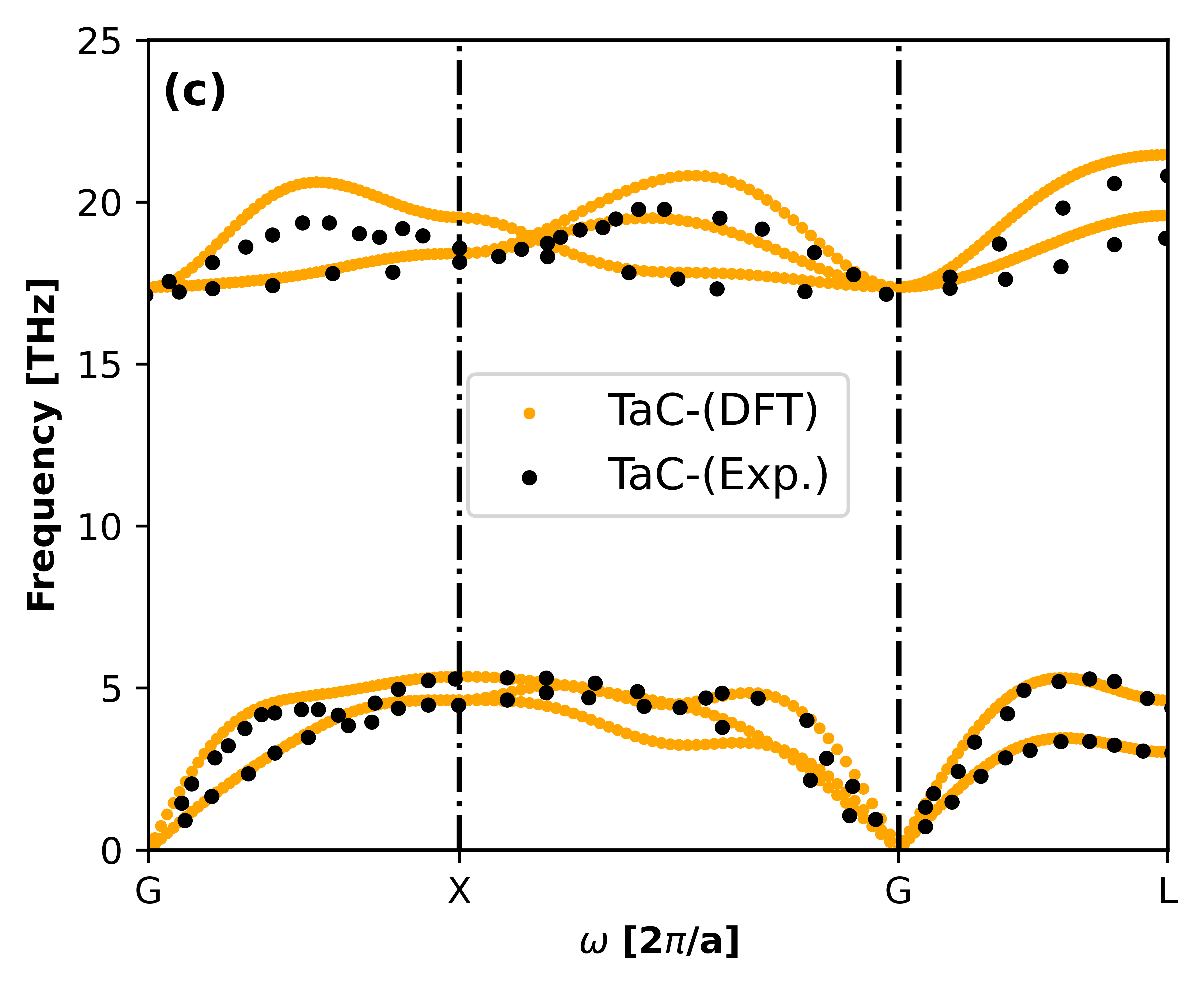}
    \end{subfigure}

    \begin{subfigure}[b]{0.3\textwidth}
        \includegraphics[width=\textwidth]{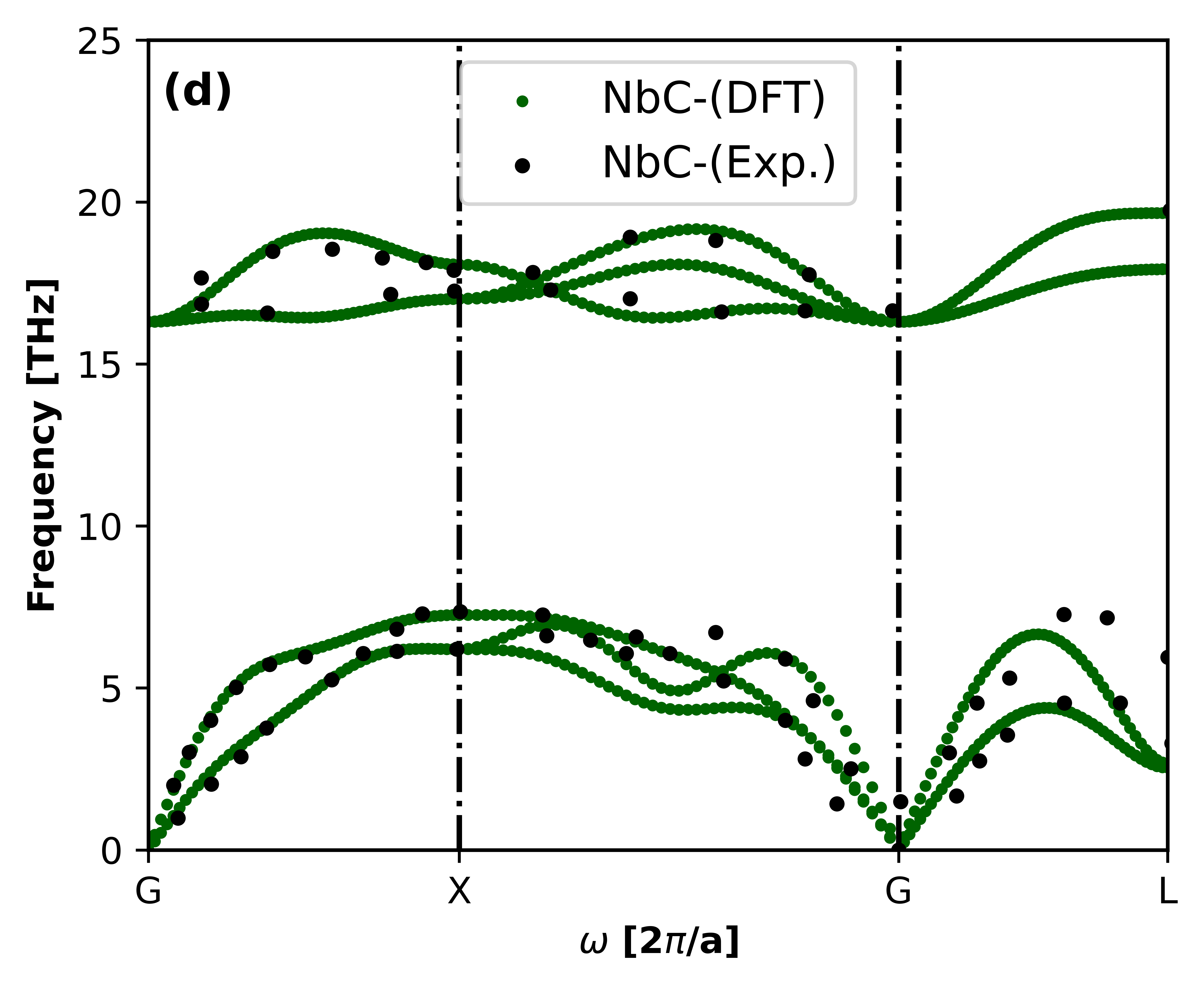}
    \end{subfigure}
    \begin{subfigure}[b]{0.3\textwidth}
        \includegraphics[width=\textwidth]{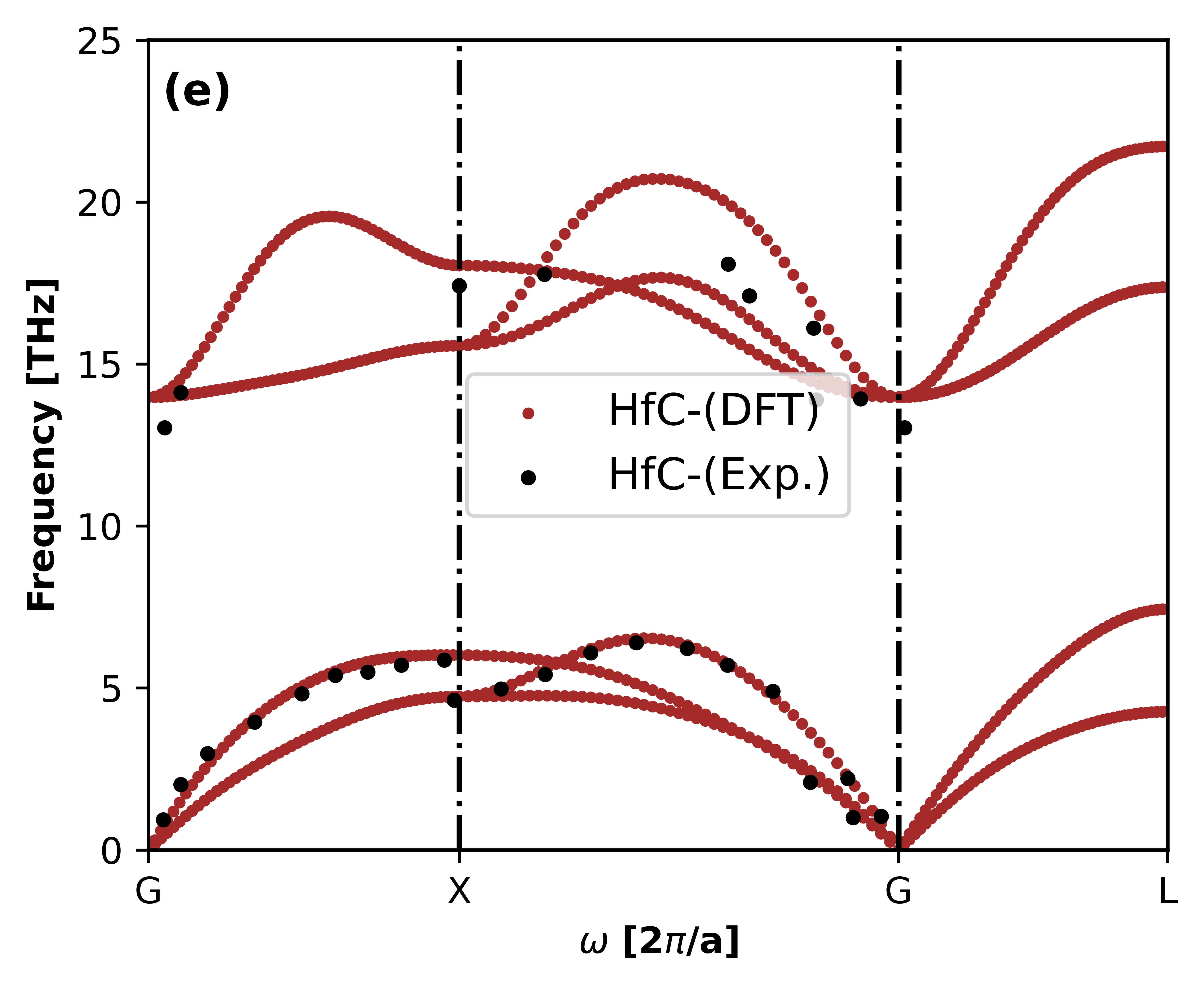}
    \end{subfigure}
    \caption{Comparison of the phonon dispersion spectra of the individual constituents of the compositionally complex carbides (a) ZrC, (b) TiC, (c) TaC, (d) NbC, (e) HfC.}
    \label{fig:1}
\end{figure*}

Overall, Figure~\ref{fig:1} presents the predicted phonon dispersion spectra of TMCs along the high-symmetry path in the Brillouin Zone ($\Gamma$-X-$\Gamma$-L). We have compared these predictions with existing experimental data from the literature and see excellent agreement \cite{smith1980_phonons_ZrC,Pintschovius_1978_TiC_exp,Smith_PRL_1972,Smith_1970_TaC_HfC}. The predicted results successfully capture key features of these TMCs, including the anomalous softening of the longitudinal acoustic (LA) modes in superconducting TMCs like TaC and NbC, the larger optical mode splitting in non-superconducting TMCs such as ZrC, TiC and HfC compared to the superconducting TMCs, and the significant bandgap between the acoustic and optical phonon branches. Among these properties, the magnitude of the phonon bandgap can significantly influence thermal conductivity \cite{Portwin2024}. In certain materials like boron arsenide \cite{Lindsay2013} and titanium hydride \cite{Wang2017}, the bandgap size plays a crucial role in modulating thermal properties. Specifically, a phonon bandgap reduces the available phase space for phonon-phonon scattering events, thereby increasing phonon lifetimes and enhancing the thermal conductivity \cite{Maldovan2015,Qian2021}. Therefore, we present the phonon bandgaps of each of the TMCs at the X-point, $\mathbf{k}=(\frac{1}{2}, 0 ,\frac{1}{2}$) to provide reference values for comparison. Figure~\ref{fig:1}(a--e) shows the phonon disperion spectra of ZrC, TiC, TaC, NbC, and HfC respectively. The corresponding X-point bandgaps are 6.27, 5.64, 13.05, 9.75, and 9.55 THz. From this analysis, we understand that the phonon bandgap at the X point increases in the following order: TiC $<$ ZrC $<$ HfC  $<$ NbC  $<$ TaC. Among these, ZrC, TiC and HfC exhibit relatively narrow bandgaps, while they are much wider in TaC and NbC. Based on these results, we can use these TMCs to modulate the phonon bandgap and therefore the thermal conductivity of the CCCs. The change in the phonon bandgap can be mainly attributed to two factors: the atomic mass and chemical nature of the transition metal ion. For elements in the same group, like Ti, Zr, and Hf, or Nb and Ta, larger atomic mass increases the phonon bandgap by reducing the energy of acoustic phonons. For nearby elements with similar masses, like Zr and Nb, or Hf and Ta, the extra electron in Nb and Ta increases the energy of optical phonons due to different bond strengths. We believe this variation could be used as a pathway to engineer the phonon bandgap and broadening by altering the composition of high entropy carbides, transitioning from binary to quinary systems. This approach could thereby tailor the thermal properties of the CCCs for advanced technological applications.

\begin{figure*}[ht!]
  \centering
  \begin{subfigure}[b]{0.32\textwidth}
    \centering
    \includegraphics[width=\textwidth]{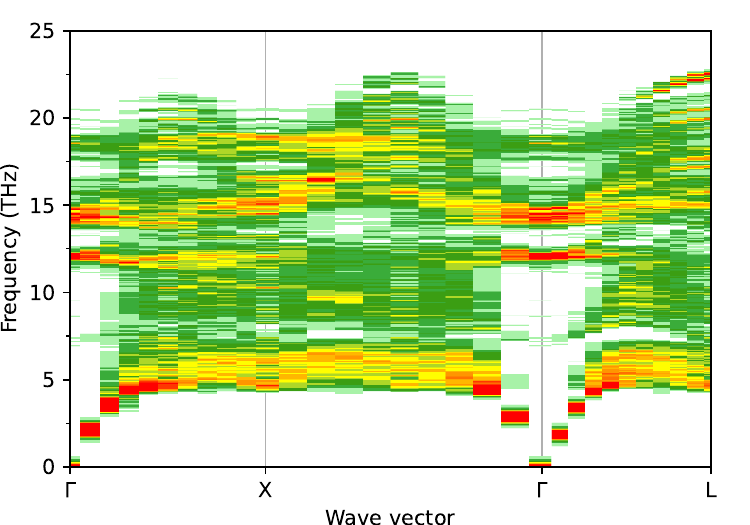}
    \caption{}
    \label{fig:2a}
  \end{subfigure}
  \hfill
  \begin{subfigure}[b]{0.32\textwidth}
    \centering
    \includegraphics[width=\textwidth]{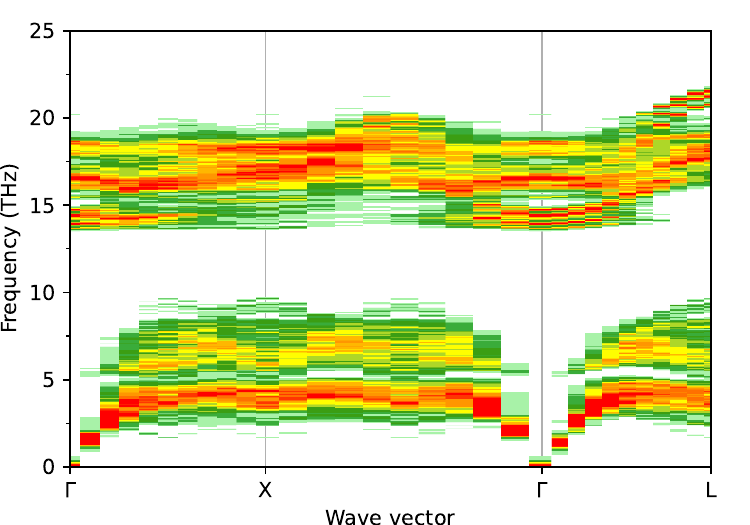}
    \caption{}
    \label{fig:2b}
  \end{subfigure}
  \hfill
  \begin{subfigure}[b]{0.32\textwidth}
    \centering
    \includegraphics[width=\textwidth]{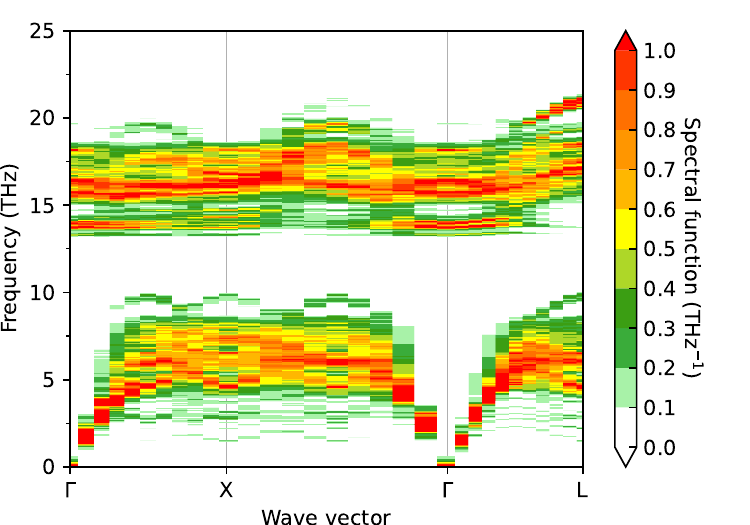}
    \caption{}
    \label{fig:2c}
  \end{subfigure}
    \hfill
    \begin{subfigure}[b]{0.32\textwidth}
    \centering
    \includegraphics[width=\textwidth]{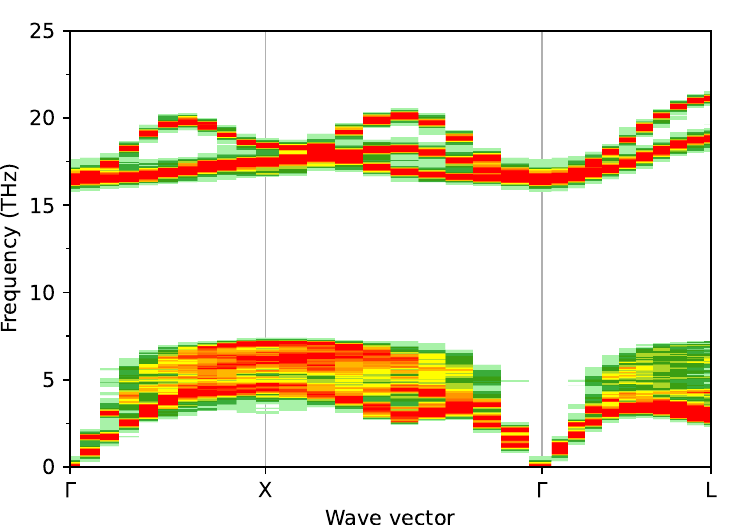}
    \caption{}
    \label{fig:2d}
  \end{subfigure}
  \begin{subfigure}[b]{0.32\textwidth}
    \centering
    \includegraphics[width=\textwidth]{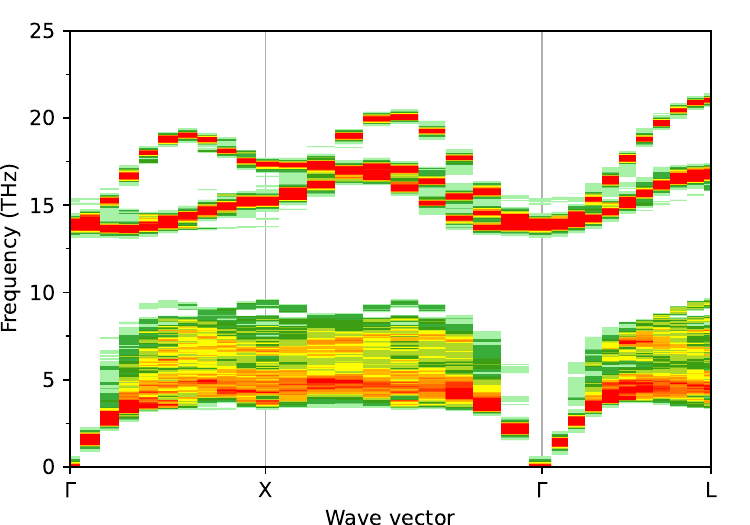}
    \caption{}
    \label{fig:2e}
  \end{subfigure}
  \hfill
  \begin{subfigure}[b]{0.32\textwidth}
    \centering
    \includegraphics[width=\textwidth]{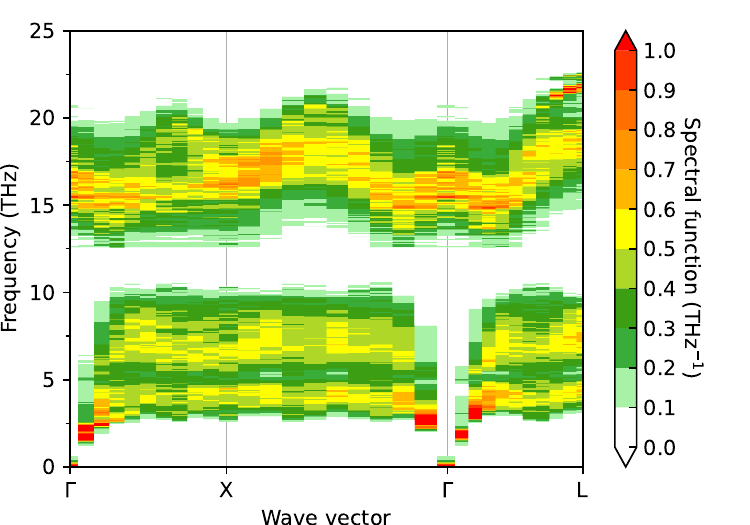}
    \caption{}
    \label{fig:2f}
  \end{subfigure}
  \begin{subfigure}[b]{0.32\textwidth}
    \centering
    \includegraphics[width=\textwidth]{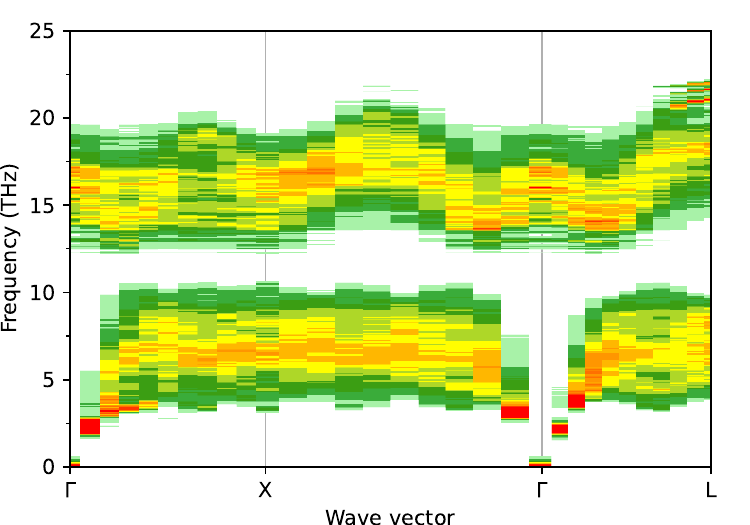}
    \caption{}
    \label{fig:2g}
  \end{subfigure}
  \hspace*{\fill} 
  \begin{subfigure}[b]{0.32\textwidth}
    \centering
    \includegraphics[width=\textwidth]{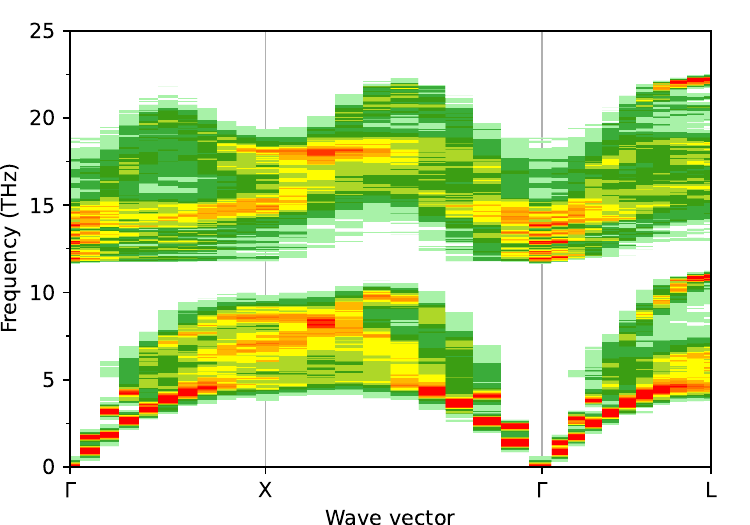}
    \caption{}
    \label{fig:2h}
  \end{subfigure}
  \hspace*{\fill} 
  \begin{subfigure}[b]{0.32\textwidth}
    \centering
    \includegraphics[width=\textwidth]{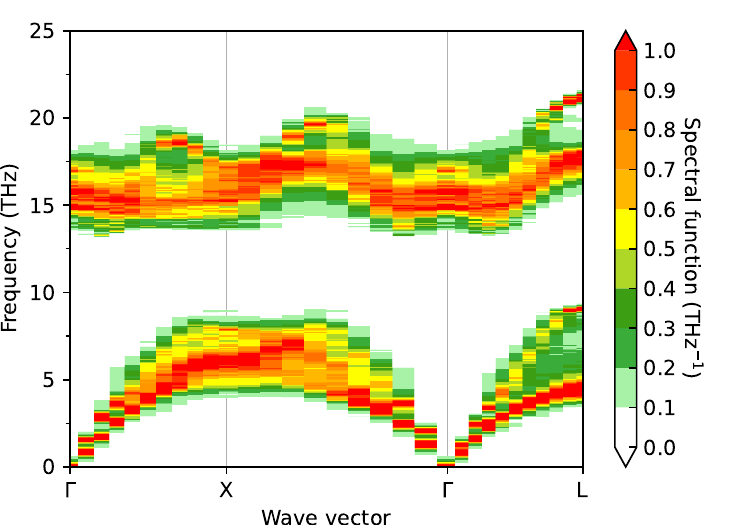}
    \caption{}
    \label{fig:2i}
  \end{subfigure}
  \hspace*{\fill}
  \begin{subfigure}[b]{0.32\textwidth}
    \centering
    \includegraphics[width=\textwidth]{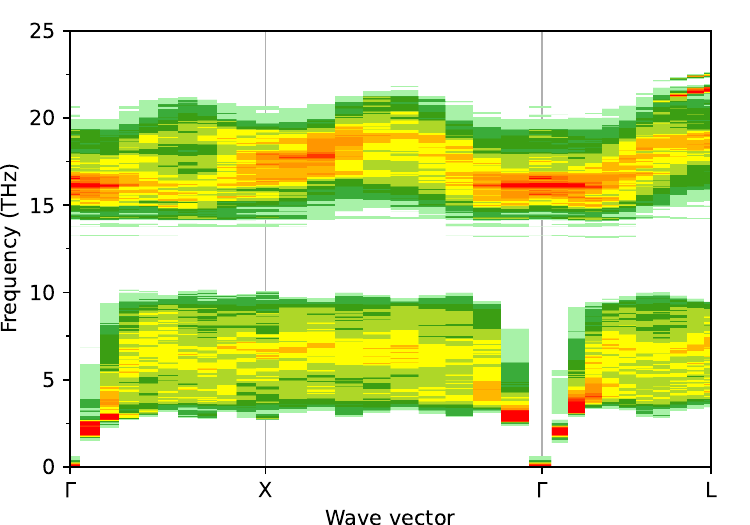}
    \caption{}
    \label{fig:2j}
  \end{subfigure}
  \hspace*{\fill}
  \begin{subfigure}[b]{0.32\textwidth}
    \centering
    \includegraphics[width=\textwidth]{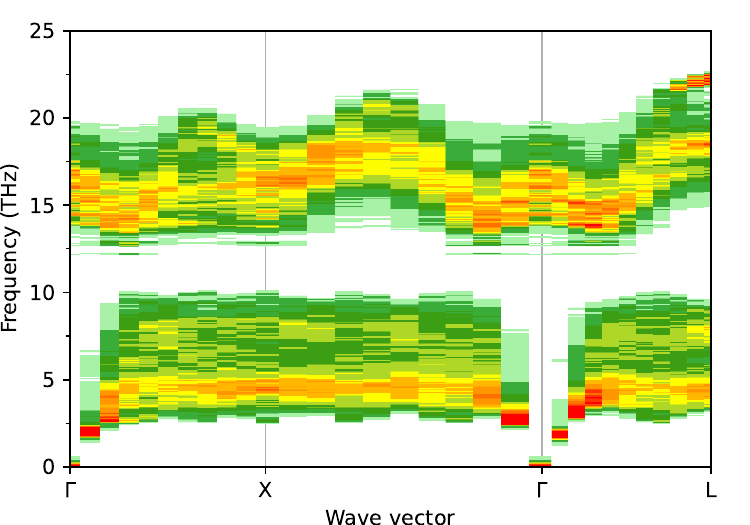}
    \caption{}
    \label{fig:2k}
  \end{subfigure}
  \hspace*{\fill}
  \begin{subfigure}[b]{0.32\textwidth}
    \centering
    \includegraphics[width=\textwidth]{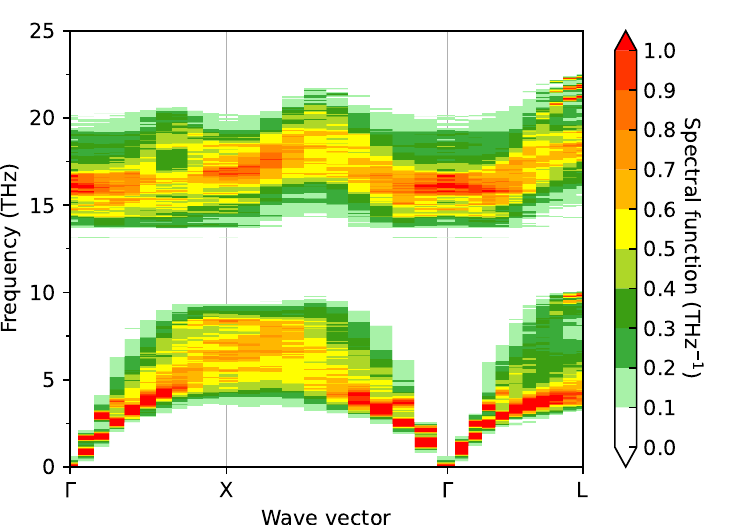}
    \caption{}
    \label{fig:2l}
  \end{subfigure}
  \caption{Comparison of the phonon dispersion spectra of binary ceramics (a) Zr$_{0.5}$Ti$_{0.5}$C, (b) Zr$_{0.5}$Ta$_{0.5}$C, (c) Zr$_{0.5}$Nb$_{0.5}$C, (d) Ta$_{0.5}$Nb$_{0.5}$C, (e) Zr$_{0.5}$Hf$_{0.5}$C, ternary ceramics (f) Zr$_{0.33}$Ti$_{0.33}$Ta$_{0.33}$C,(g) Zr$_{0.33}$Ti$_{0.33}$Nb$_{0.33}$C, (h) Zr$_{0.33}$Ti$_{0.33}$Hf$_{0.33}$C, (i)Zr$_{0.33}$Nb$_{0.33}$Hf$_{0.33}$C, quaternary ceramics (j) Zr$_{0.25}$Ti$_{0.25}$Ta$_{0.25}$Nb$_{0.25}$C, (k) Zr$_{0.25}$Ti$_{0.25}$Ta$_{0.25}$Hf$_{0.25}$C, quinary (l) ceramics Zr$_{0.2}$Ti$_{0.2}$Ta$_{0.2}$Hf$_{0.2}$Nb$_{0.2}$C.}
\label{fig2}
\end{figure*}

Figure~\ref{fig2} presents the computed phonon spectra along the $\Gamma$-X-$\Gamma$-L path for binary, ternary, quaternary, and quinary CCCs. By systematically adding different elements, we have evaluated the impact of each element on the phonon bandgap and broadening. Detailed phonon contributions of individual elements are provided in the supplementary information (SI), with the full spectra in the first column and their individually resolved contribution in the corresponding rows. For all ceramics considered, no broadening is observed in the region near the $\Gamma$ point, $\mathbf{k}=(0, 0 ,0)$, adhering to the Rayleigh criteria $(\omega^4)$ for impurity scattering \cite{FRITZ_2017}. However, significant broadening occur in the mid and high-frequency range, spanning a few THz. Another intriguing observation is that the bandgap between the acoustic and optic modes can be tuned depending on the alloying elements in the CCCs (see Figure \ref{fig:S1_overall}). For instance, in Zr$_{0.5}$Ti$_{0.5}$C (Figure~\ref{fig:2a}), the acoustic and optic modes interact, whereas other binary ceramics exhibit a phonon bandgap that is strongly dependent on the alloying elements. By decomposing the full phonon spectra into contributions from individual chemical elements, we found that in Zr$_{0.5}$Ti$_{0.5}$C, the lowest frequency phonons are from Zr, the mid range frequencies from Ti, and the highest frequencies from C atoms. In Zr$_{0.5}$Ta$_{0.5}$C (Figure~\ref{fig:2b}), the phonon frequencies up to 5 THz are from the Ta atom, frequencies above 5 to 10 THz are from Zr atom and the highest frequency from C atom, with a significant bandgap. Additionally, the phonon broadening due to C was relatively smaller in Zr$_{0.5}$Ta$_{0.5}$C compared to the Zr$_{0.5}$Ti$_{0.5}$C, which could be attributed to the relatively smaller splitting of the optical branches in TaC compared to the TiC and ZrC. In Zr$_{0.5}$Nb$_{0.5}$C (Figure~\ref{fig:2c}), Zr and Nb have similar atomic masses, and the individual contributions to the phonon frequencies are predicted in a similar range, leading to the absence of the slight gap observed around 7 THz in Zr$_{0.5}$Ti$_{0.5}$C and 5 THz in Zr$_{0.5}$Ta$_{0.5}$C. Among all the binary ceramics the bandgap between the acoustic and the optic phonon in the Ta$_{0.5}$Nb$_{0.5}$C (Figure~\ref{fig:2d}) is the largest, consistent with the large bandgap of the constituent TaC and NbC. Similarly, the binary alloy of Zr$_{0.5}$Hf$_{0.5}$C (Figure~\ref{fig:2e}) also shows a significant phonon bandgap, where Hf contributes to the modes at lower frequency up to 5THz. From the comparison of the phonon spectra of the different binary ceramics, we can clearly see that Zr$_{0.5}$Ti$_{0.5}$C has no phonon bandgap and the largest phonon broadening, while Ta$_{0.5}$Nb$_{0.5}$C has the largest phonon bandgap and the least phonon broadening. Intuitively, we can infer that the phonon thermal conductivity will be significantly reduced in Zr$_{0.5}$Ti$_{0.5}$C ceramics due to the smaller scattering phase space available, compared to that of Ta$_{0.5}$Nb$_{0.5}$C.  

In our investigation of ternary ceramics, we considered four compositions: Zr$_{0.33}$Ti$_{0.33}$Ta$_{0.33}$C (Figure \ref{fig:2f}), Zr$_{0.33}$Ti$_{0.33}$Nb$_{0.33}$C (Figure \ref{fig:2g}), Zr$_{0.33}$Ti$_{0.33}$Hf$_{0.33}$C (Figure \ref{fig:2h}), and Zr$_{0.33}$Nb$_{0.33}$Hf$_{0.33}$C (Figure \ref{fig:2i}). In the binary composition Zr$_{0.5}$Ti$_{0.5}$C (Figure~\ref{fig:2a}), no phonon bandgap was observed between the acoustic and optic phonons. However, the introduction of TaC into Zr$_{0.5}$Ti$_{0.5}$C, resulted in a phonon bandgap in the ternary alloy Zr$_{0.33}$Ti$_{0.33}$Ta$_{0.33}$C (Figure~\ref{fig:2f}), indicating that phonon bandgaps can be tuned in CCCs. Analyzing the contribution from individual chemical elements (Figure \ref{fig:S2_overall}), we note that Ti, which was responsible for the absence of a bandgap in the binary composition Zr$_{0.5}$Ti$_{0.5}$C, showed a weaker contribution in Zr$_{0.33}$Ti$_{0.33}$Ta$_{0.33}$C due to its reduced concentration in the alloy. This behavior is consistent across all Zr$_{0.33}$Ti$_{0.33}$M$_{0.33}$C ternary ceramics (M=Ta, Nb, Hf). The equiatomic addition of TaC, NbC, or HfC to Zr$_{0.5}$Ti$_{0.5}$C results in the formation of a bandgap, with the size of the bandgap being largest for TaC, followed by NbC and then HfC. These findings align with the work by Körmann et al. \cite{FRITZ_2017}, who reported a weaker contribution of each element in the phonon dispersion spectra due to reduced element concentration in the high entropy alloy (MoTaNbWV). Similarly, we find that decreasing the relative concentration of both Zr and Ti allows a bandgap to form, with the size of the bandgap determined by the specific TMC being added. Another interesting observation is that, when Zr$_{0.5}$Nb$_{0.5}$C is used as a base, and HfC is added, the broadening of the phonon dispersion spectra is reduced, resulting in an increased bandgap. This suggests that Zr$_{0.33}$Nb$_{0.33}$Hf$_{0.33}$C may be a better carrier of heat through phonons compared to the other ternary compositions considered in this work. The phonon band structures of two quaternary compositions, Zr$_{0.25}$Ti$_{0.25}$Ta$_{0.25}$Nb$_{0.25}$C (Figure~\ref{fig:2j}) and Zr$_{0.25}$Ti$_{0.25}$Ta$_{0.25}$Hf$_{0.25}$C (Figure~\ref{fig:2k}), showed similar characteristics. However, the spectral function reveals a higher phonon density at lower frequencies for Zr$_{0.25}$Ti$_{0.25}$Ta$_{0.25}$Hf$_{0.25}$C compared to Zr$_{0.25}$Ti$_{0.25}$Ta$_{0.25}$Nb$_{0.25}$C (Figure \ref{fig:S3_overall} provides detailed contribution from individual elements). 

Figure~\ref{fig:2k} presents the predicted phonon band structure of the quinary ceramic Zr$_{0.2}$Ti$_{0.2}$Ta$_{0.2}$Hf$_{0.2}$Nb$_{0.2}$C. It was intriguing to note that the bandgap in the quinary composition is more pronounced compared to the quaternary ceramics, and some binary and ternary compositions. Additionally, the phonon broadening is reduced in the quinary composition relative to the quaternary, and some of the binary and ternary compositions. This indicates that phonon transport in the quinary composition could be superior to that in the quaternary and potentially even some ternary and binary  compositions. Overall, this study demonstrates that the phonon bandgap and the phonon spectral function, and therefore thermal conductivity, can be tuned based on the specific elemental combinations and the concentration of the components involved.

\begin{figure}[h]
    \centering
    \includegraphics[width=0.45\textwidth]{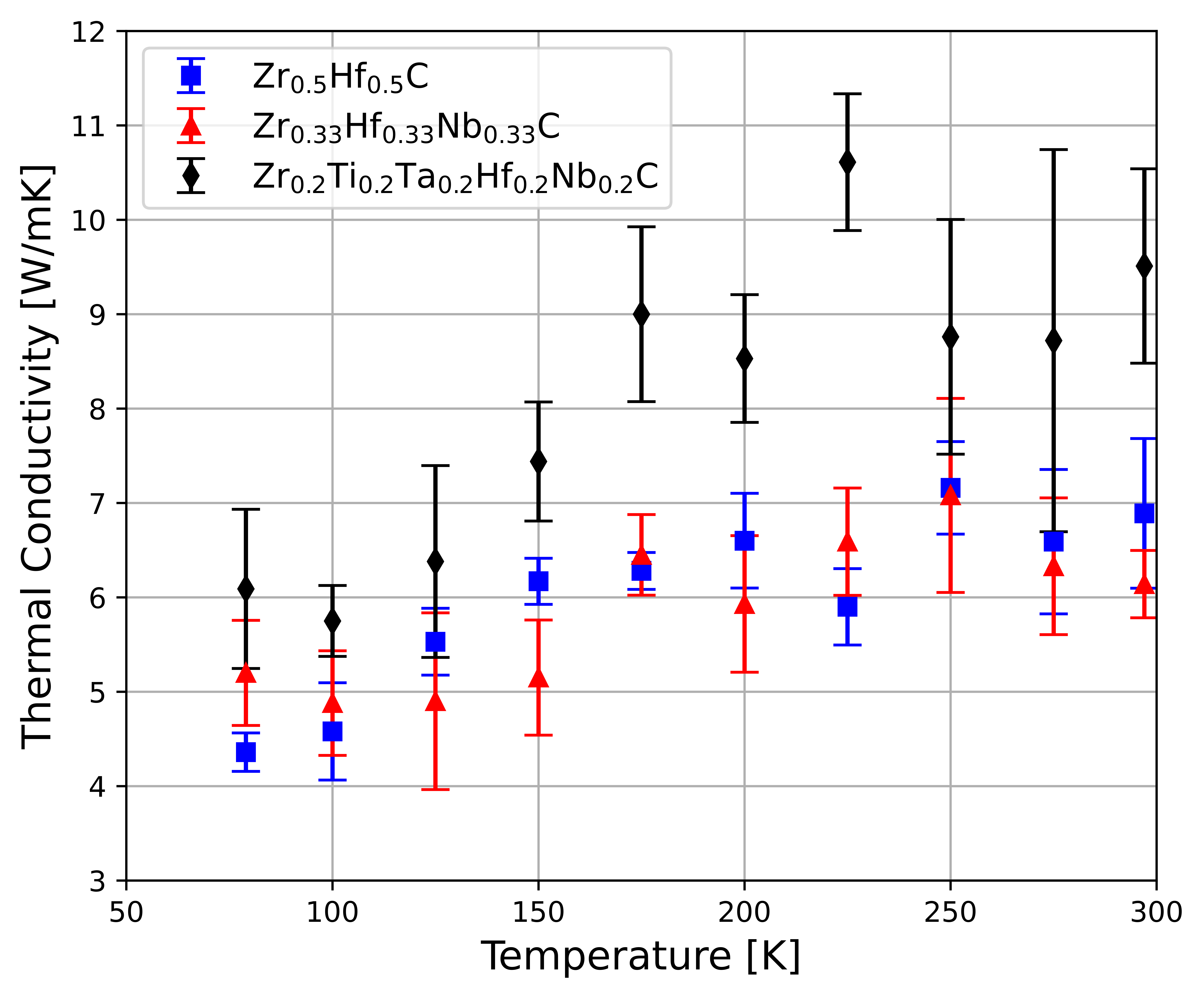}
    \caption{Thermal conductivity of the quinary composition (Zr$_{0.2}$Ti$_{0.2}$Ta$_{0.2}$Hf$_{0.2}$Nb$_{0.2}$C) along with a binary (Zr$_{0.5}$Hf$_{0.5}$C) and ternary ceramics  (Zr$_{0.33}$Hf$_{0.33}$Nb$_{0.33}$C) at low temperature.}
    \label{fig:3}
\end{figure}

Since predicting the thermal conductivity of these CCCs was beyond the scope of this work, we report the measured thermal conductivity of the quinary CCC (Zr$_{0.2}$Ti$_{0.2}$Ta$_{0.2}$Hf$_{0.2}$Nb$_{0.2}$C) along with a binary (Zr$_{0.5}$Hf$_{0.5}$C) and ternary ceramic carbide (Zr$_{0.33}$Hf$_{0.33}$Nb$_{0.33}$C) for comparison. Figure~\ref{fig:3} shows the thermal conductivity of these three carbides measured at low temperatures using the SDTR technique, where the phonon contribution is a major factor in the total thermal conductivity. The results clearly show that the quinary CCC has higher thermal conductivity compared to the measured binary and ternary carbides. This aligns with the observations above, suggesting that the quinary CCC would have superior thermal conductivity compared to some of the binary and ternary carbides. However, there are several factors that could contribute to this observation, such as a wider phonon bandgap and reduced phonon broadening in the quinary CCC, therefore, we do not explicitly identify the mechanisms by which the thermal conductivity in the quinary CCC is superior. This remains an open question that requires further investigation and development to predict the thermal conductivity of complex systems. However, this work lays a foundation for exploring materials with better thermal conductivity than some of the existing high entropy materials. 

\section{Conclusion}
Using a fully \textit{ab initio} method, we have presented an extensive analysis of the phonon-band structure and phonon broadening in CCCs. The results reveal that the phonon band structure of the CCCs depends on the elements and their concentrations in the ceramics. The study also indicates that the phonon bandgap between the acoustic and optic modes can be tuned based on the elements and their relative concentrations. Generally, the heaviest atom contributes to the lowest frequency phonons, while the lightest atom contributes to the highest frequency phonons. Thus, by varying the elements with different atomic masses, one can tune the phonon frequencies of the CCCs, thereby controlling various physical properties. Additionally, we report the measured thermal conductivity of the quinary ceramic, along with binary and ternary ceramics. The data indicates that the quinary CCC may exhibit superior thermal properties, which is unexpected given the anticipated increase in scattering and reduction in thermal conductivity due to greater cation disorder. These results suggest that certain CCCs can achieve enhanced thermal properties. The predictions of the phonon dispersion spectra, coupled with the measured thermal conductivities of the CCCs, provide a foundational understanding for exploring an intriguing design parameter for thermal conductivity. These findings will assist researchers in developing new CCC compositions for potential applications in extreme environments, such as carbide fuels or claddings in advanced fission reactors, coatings for fusion reactor components, and leading edges of hypersonic vehicles.

\section{Acknowledgment}
This work was supported by the US Department of Energy, Office of Nuclear Energy under DOE Idaho Operations Office Contract DE-AC07-05ID14517 and LDRD project of 22A1059-107FP. The authors also acknowledge that this research made use of the resources of the High Performance Computing Center at Idaho National Laboratory, which is supported by the Office of Nuclear Energy of the U.S. Department of Energy and the Nuclear Science User Facilities under Contract No. DE-AC07-05ID14517. The United States Government retains and the publisher, by accepting the article for publication, acknowledges that the U.S. Government retains a nonexclusive, paid-up, irrevocable, worldwide license to publish or reproduce the published form of this manuscript, or allow others to do so, for U.S. Government purposes.

\section{Additional Information}
\textbf{Supplementary information} accompanies the paper on the journal website; The SI contains the figures regarding the elemental contribution to the total phonon dispersion spectra for all the ceramics considered in this paper. 

\textbf{Competing Interests}: The authors declare that they have no competing financial interests. 

\bibliography{apssamp}

\clearpage

\onecolumngrid

\section*{Supplementary Information}
\begin{center}
    \textbf{\large Engineering Phonon bandgaps in Compositionally Complex Ceramic Carbides}
    
    \vspace{0.5em}
    
  \vspace{0.5em}
    
    Linu Malakkal\textsuperscript{1,*}, Jarin C. French\textsuperscript{1}, Kaustubh K. Bawane\textsuperscript{1}, Shuxiang Zhou\textsuperscript{1}, Zilong Hua\textsuperscript{1}, Lingfeng He\textsuperscript{2}, Yongfeng Lu \textsuperscript{3}, Bai Cui \textsuperscript{3}
    
    \vspace{0.5em}
    
    \textsuperscript{1}Idaho National Laboratory, Idaho Falls, ID 83415
    \textsuperscript{2}North Carolina State University, Rayleigh
    \textsuperscript{3}University of Nebaraska Lincoln, 1400 R St, Lincoln, NE 68588
    \vspace{1em}
    
    \textsuperscript{*}Corresponding authors: Linu Malakkal (linu.malakkal@inl.gov)
    
    \vspace{1em}
    
    \today
\end{center}
\section*{Decomposing the Full Phonon Spectra into Contributions from Individual Chemical Elements}
In the supplementary section, we provide the details of the individual element contribution to the total phonon dispersion spectra of all the different compositionally complex carbides (CCCs) considered in this work. 

\clearpage
\renewcommand{\thefigure}{S\arabic{figure}}
\setcounter{figure}{0}
\noindent\begin{minipage}{\textwidth}
    Binary ceramics:
\end{minipage}
\vspace{1em} 
\begin{figure*}[ht!]
  \centering
  \begin{subfigure}[b]{0.24\textwidth}
    \centering
    \includegraphics[width=\textwidth]{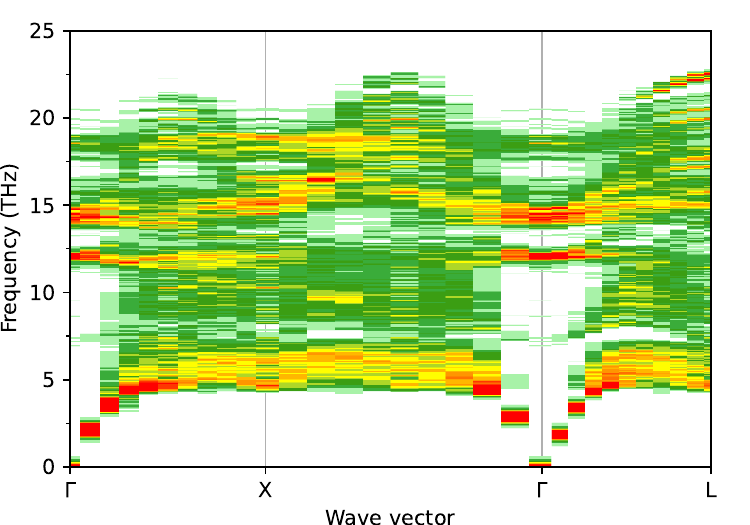}
    \caption{Zr$_{0.5}$Ti$_{0.5}$C}
    \label{fig:S1a}
  \end{subfigure}
  \hfill
  \begin{subfigure}[b]{0.24\textwidth}
    \centering
    \includegraphics[width=\textwidth]{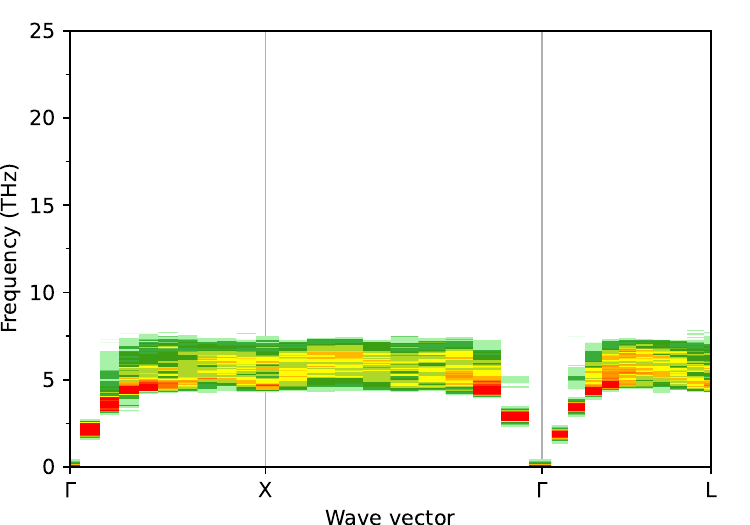}
    \caption{Zr}
    \label{fig:S1b}
  \end{subfigure}
  \hfill
  \begin{subfigure}[b]{0.24\textwidth}
    \centering
    \includegraphics[width=\textwidth]{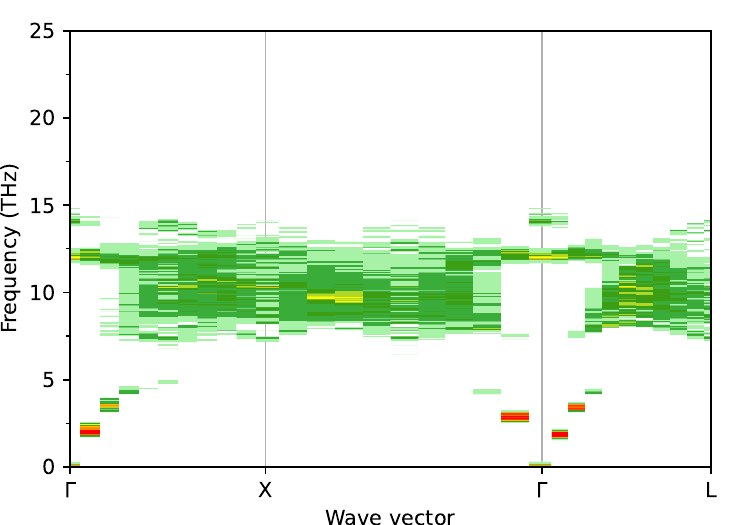}
    \caption{Ti}
    \label{fig:S1c}
  \end{subfigure}
  \hfill
  \begin{subfigure}[b]{0.24\textwidth}
    \centering
    \includegraphics[width=\textwidth]{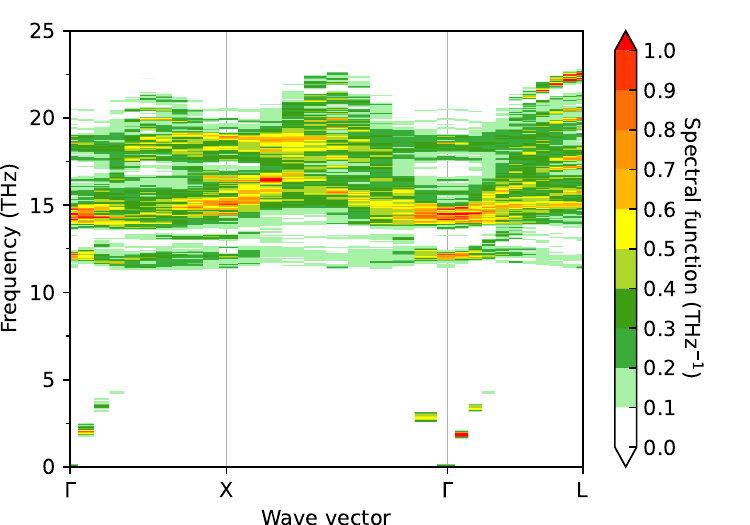}
    \caption{C}
    \label{fig:S1d}
  \end{subfigure}
  \centering
  \begin{subfigure}[b]{0.24\textwidth}
    \centering
    \includegraphics[width=\textwidth]{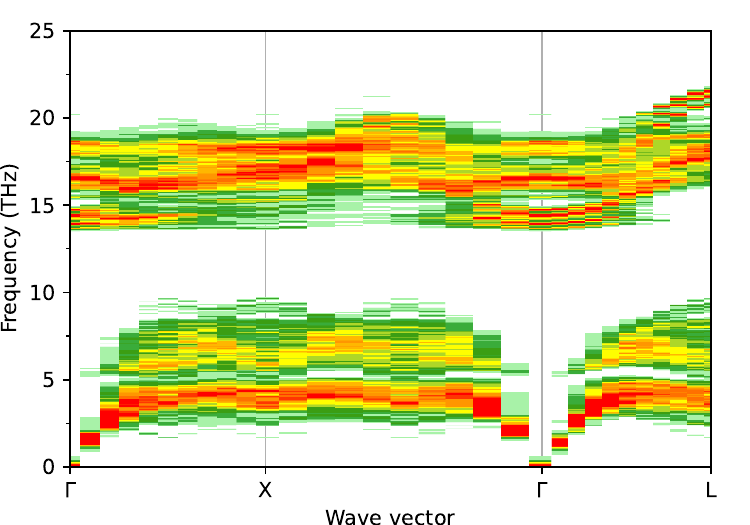}
    \caption{Zr$_{0.5}$Ta$_{0.5}$C}
    \label{fig:S1e}
  \end{subfigure}
  \hfill
  \begin{subfigure}[b]{0.24\textwidth}
    \centering
    \includegraphics[width=\textwidth]{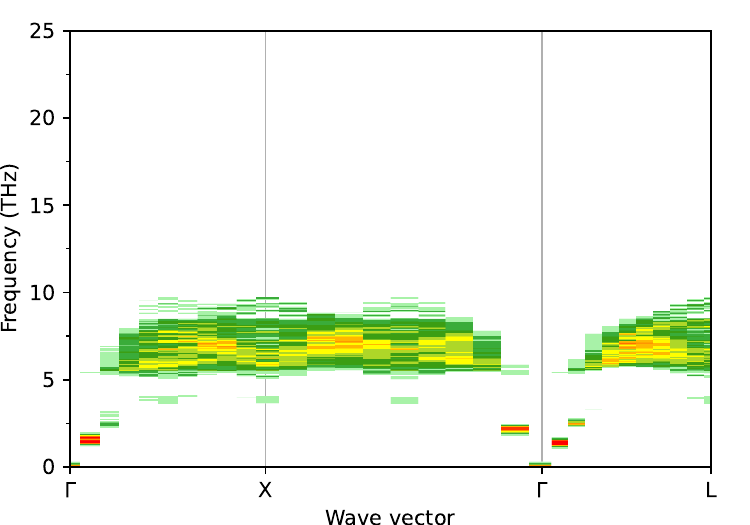}
    \caption{Zr}
    \label{fig:S1f}
  \end{subfigure}
  \hfill
  \begin{subfigure}[b]{0.24\textwidth}
    \centering
    \includegraphics[width=\textwidth]{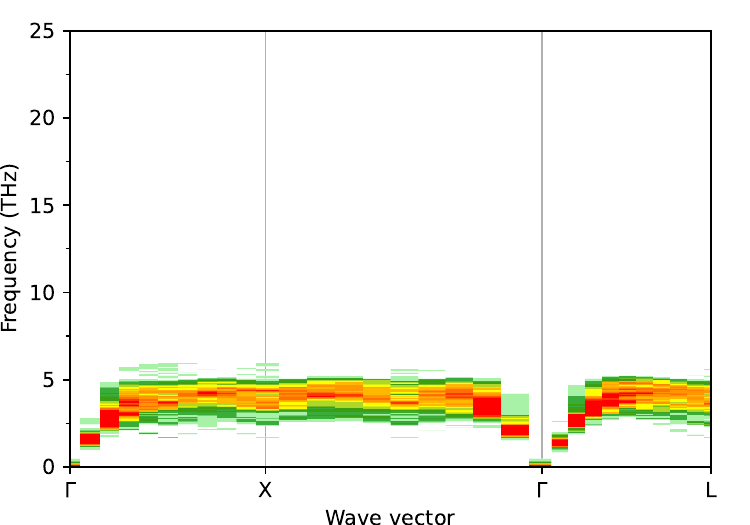}
    \caption{Ta}
    \label{fig:S1g}
  \end{subfigure}
  \hfill
  \begin{subfigure}[b]{0.24\textwidth}
    \centering
    \includegraphics[width=\textwidth]{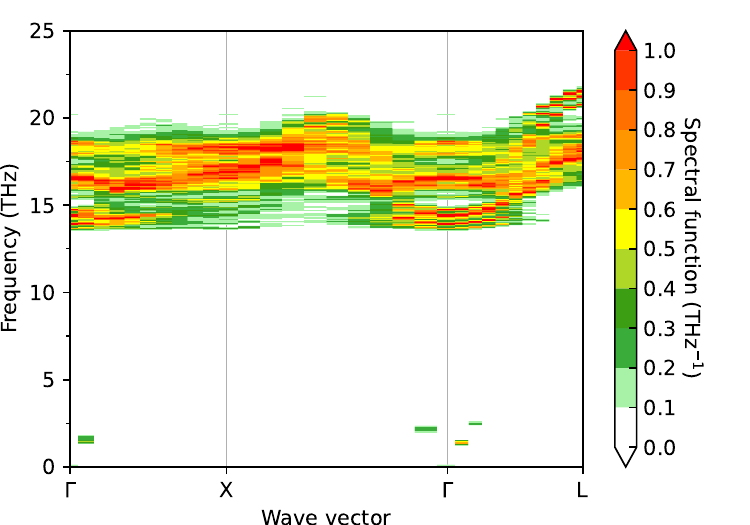}
    \caption{C}
    \label{fig:S1h}
  \end{subfigure}
  \centering
  \begin{subfigure}[b]{0.24\textwidth}
    \centering
    \includegraphics[width=\textwidth]{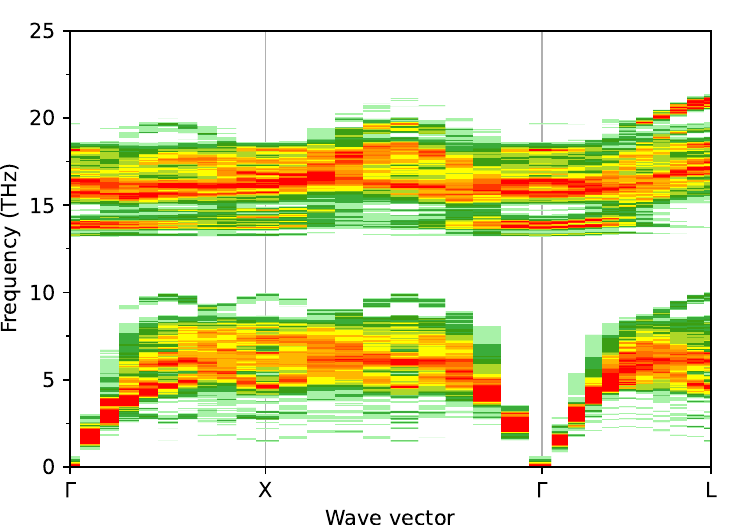}
    \caption{Zr$_{0.5}$Nb$_{0.5}$C}
    \label{fig:S1i}
  \end{subfigure}
  \hfill
  \begin{subfigure}[b]{0.24\textwidth}
    \centering
    \includegraphics[width=\textwidth]{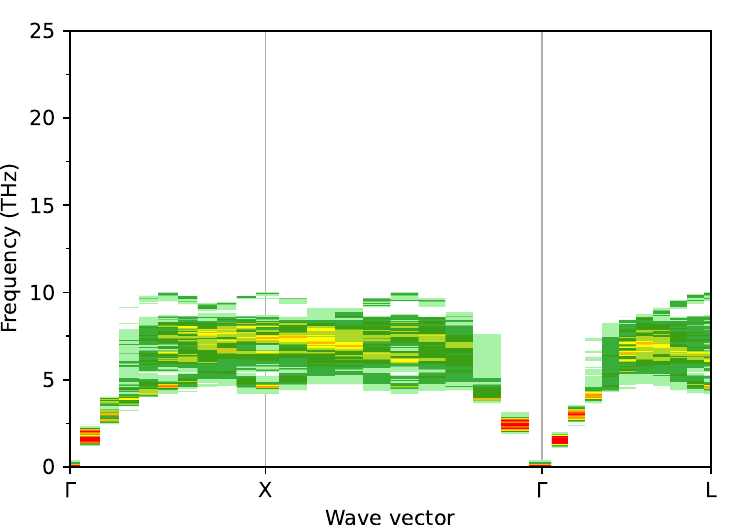}
    \caption{Zr}
    \label{fig:S1j}
  \end{subfigure}
  \hfill
  \begin{subfigure}[b]{0.24\textwidth}
    \centering
    \includegraphics[width=\textwidth]{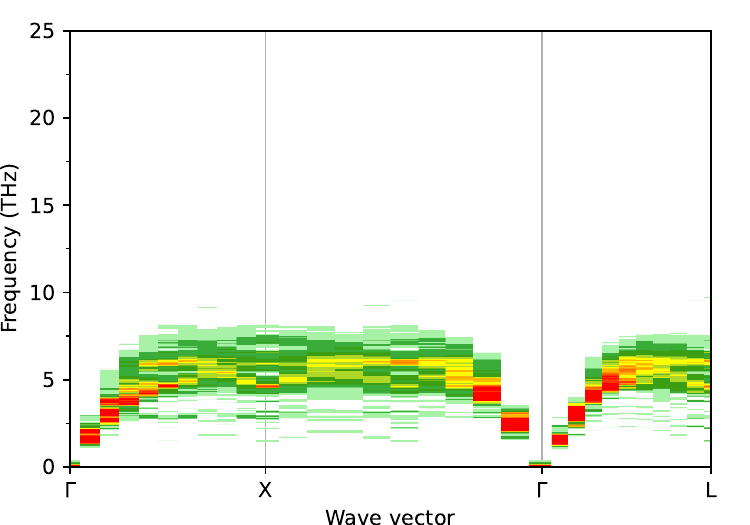}
    \caption{Nb}
    \label{fig:S1k}
  \end{subfigure}
  \hfill
  \begin{subfigure}[b]{0.24\textwidth}
    \centering
    \includegraphics[width=\textwidth]{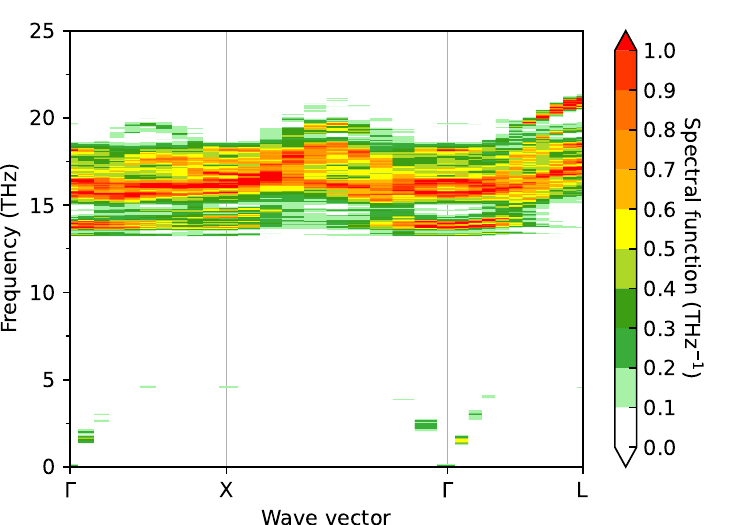}
    \caption{C}
    \label{fig:S1l}
  \end{subfigure}
  \centering
  \begin{subfigure}[b]{0.24\textwidth}
    \centering
    \includegraphics[width=\textwidth]{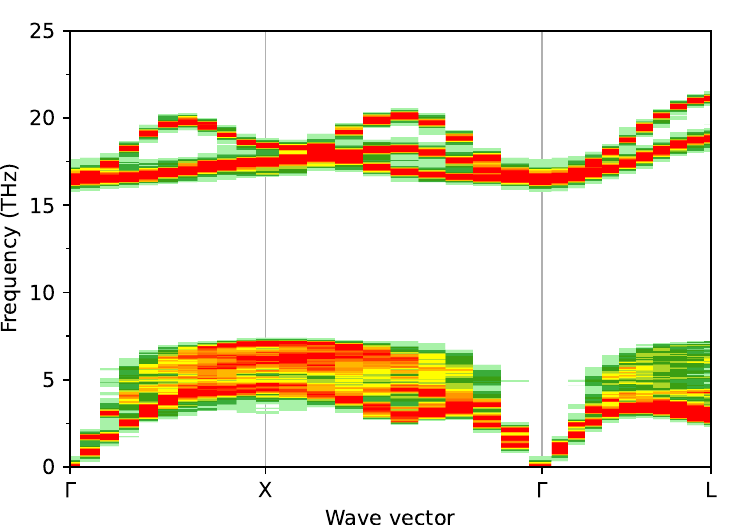}
    \caption{Ta$_{0.5}$Nb$_{0.5}$C}
    \label{fig:S1m}
  \end{subfigure}
  \hfill
  \begin{subfigure}[b]{0.24\textwidth}
    \centering
    \includegraphics[width=\textwidth]{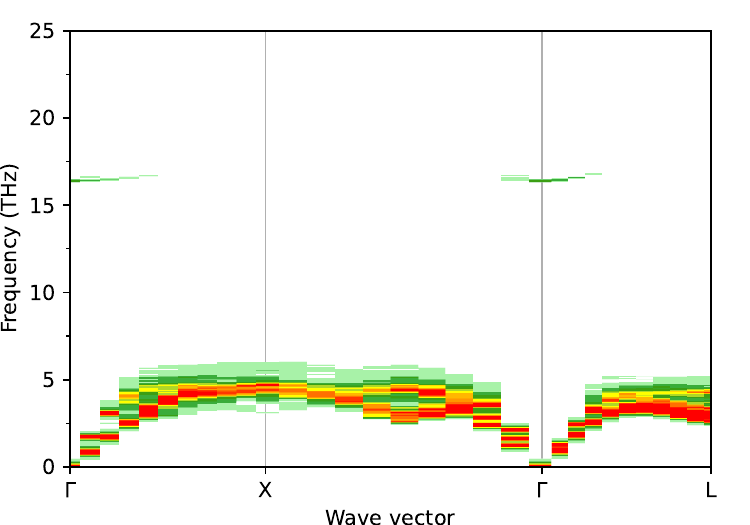}
    \caption{Ta}
    \label{fig:S1n}
  \end{subfigure}
  \hfill
  \begin{subfigure}[b]{0.24\textwidth}
    \centering
    \includegraphics[width=\textwidth]{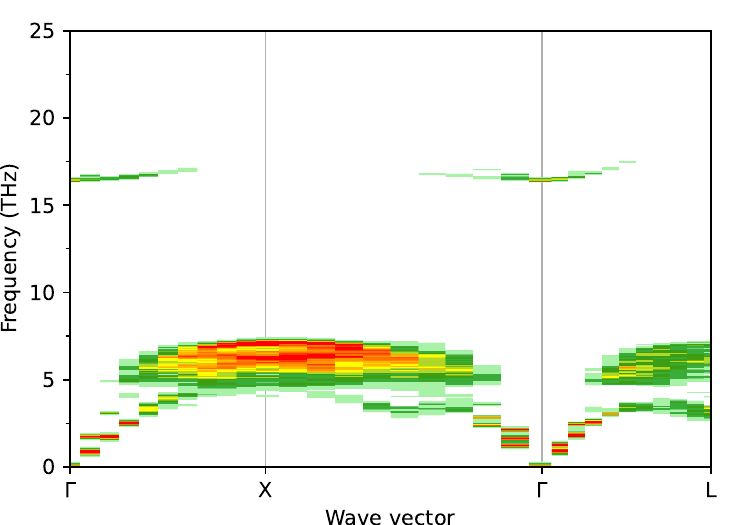}
    \caption{Nb}
    \label{fig:S1o}
  \end{subfigure}
  \hfill
  \begin{subfigure}[b]{0.24\textwidth}
    \centering
    \includegraphics[width=\textwidth]{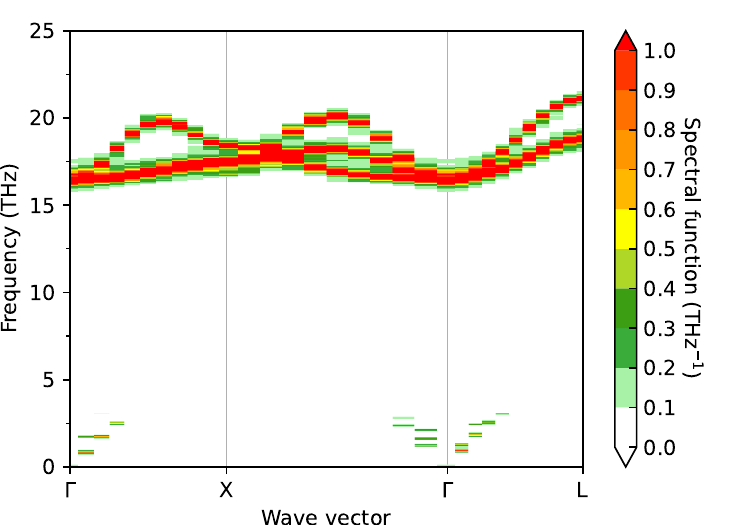}
    \caption{C}
    \label{fig:S1p}
  \end{subfigure}
  \centering
  \begin{subfigure}[b]{0.24\textwidth}
    \centering
    \includegraphics[width=\textwidth]{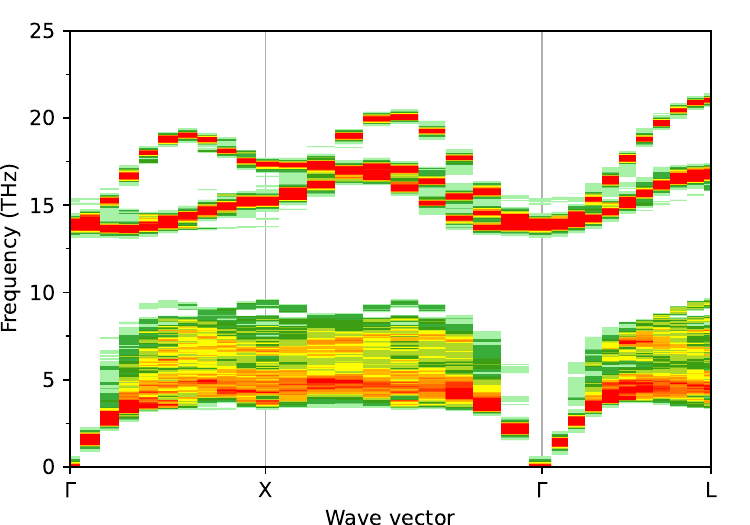}
    \caption{Zr$_{0.5}$Hf$_{0.5}$C}
    \label{fig:S1q}
  \end{subfigure}
  \hfill
  \begin{subfigure}[b]{0.24\textwidth}
    \centering
    \includegraphics[width=\textwidth]{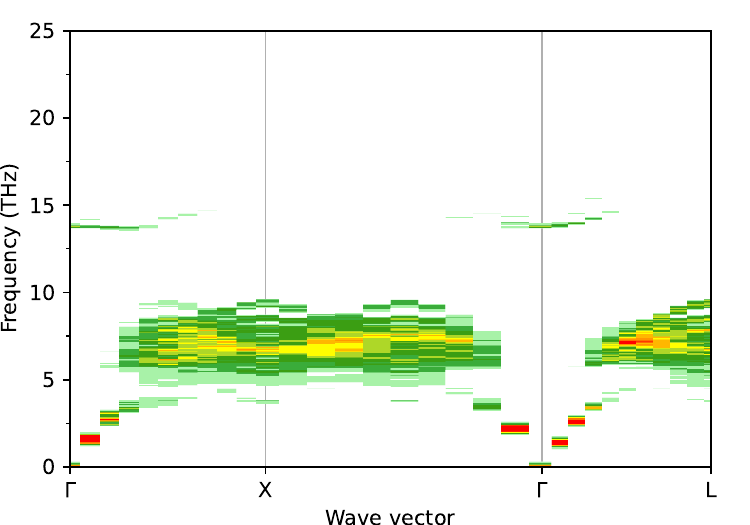}
    \caption{Zr}
    \label{fig:S1r}
  \end{subfigure}
  \hfill
  \begin{subfigure}[b]{0.24\textwidth}
    \centering
    \includegraphics[width=\textwidth]{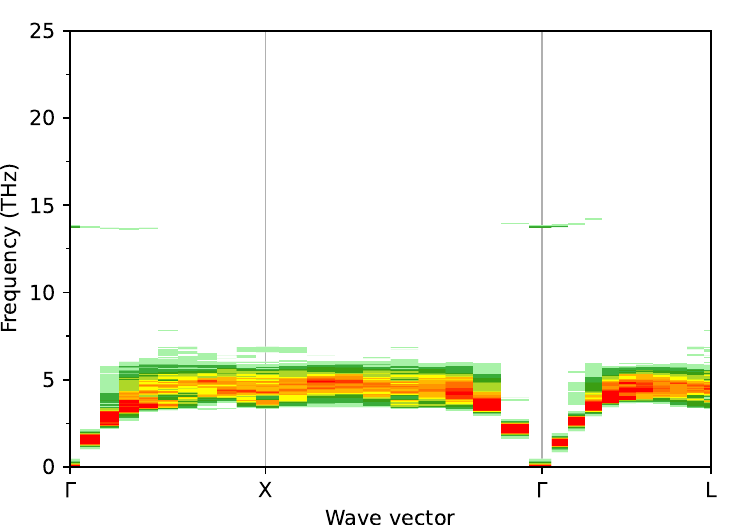}
    \caption{Hf}
    \label{fig:S1s}
  \end{subfigure}
  \hfill
  \begin{subfigure}[b]{0.24\textwidth}
    \centering
    \includegraphics[width=\textwidth]{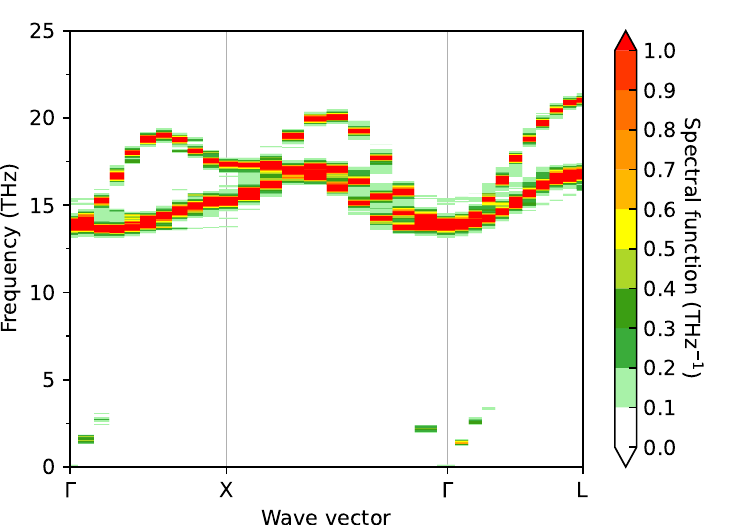}
    \caption{C}
    \label{fig:S1t}
  \end{subfigure}
\caption{\label{fig:S1_overall}Decomposition of the full phonon spectra of the binary ceramics into the individual elements. The first column is the full phonon spectra and the other columns are the contributions from the individual elements.}
\end{figure*}

\clearpage
Ternary ceramics:
\begin{figure*}[ht!]
  \centering
  \begin{subfigure}[b]{0.19\textwidth}
    \centering
    \includegraphics[width=\textwidth]{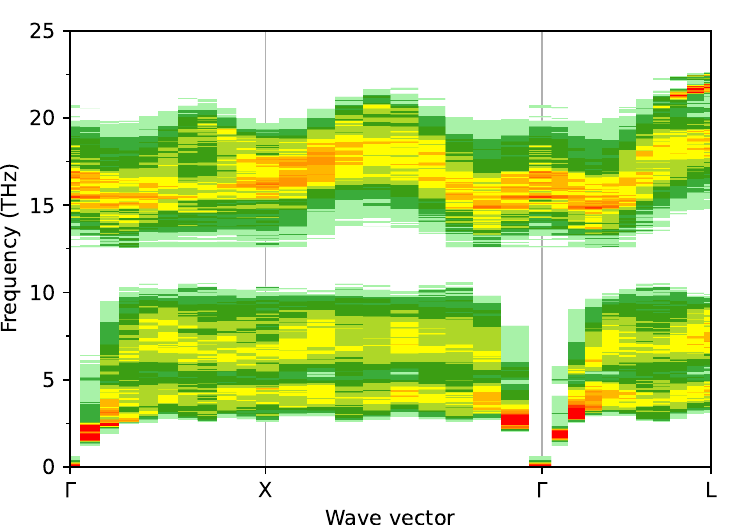}
    \caption{Zr$_{0.33}$Ti$_{0.33}$Ta$_{0.33}$C}
    \label{fig:S2a}
  \end{subfigure}
  \hfill
  \begin{subfigure}[b]{0.19\textwidth}
    \centering
    \includegraphics[width=\textwidth]{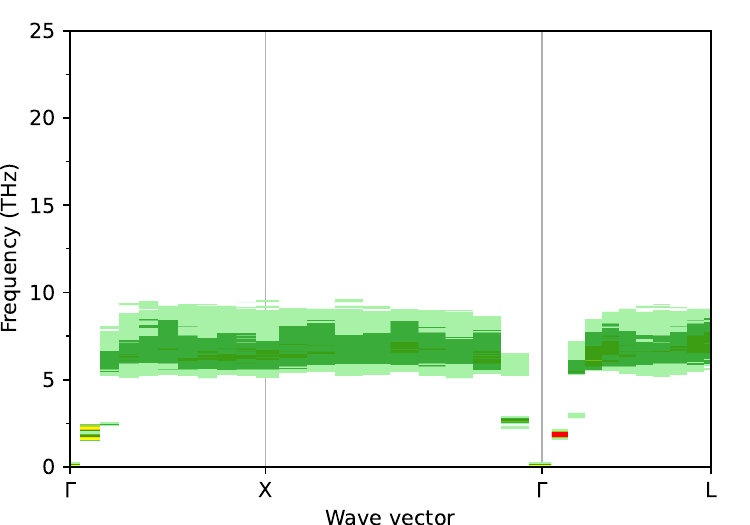}
    \caption{Zr}
    \label{fig:S2b}
  \end{subfigure}
  \hfill
  \begin{subfigure}[b]{0.19\textwidth}
    \centering
    \includegraphics[width=\textwidth]{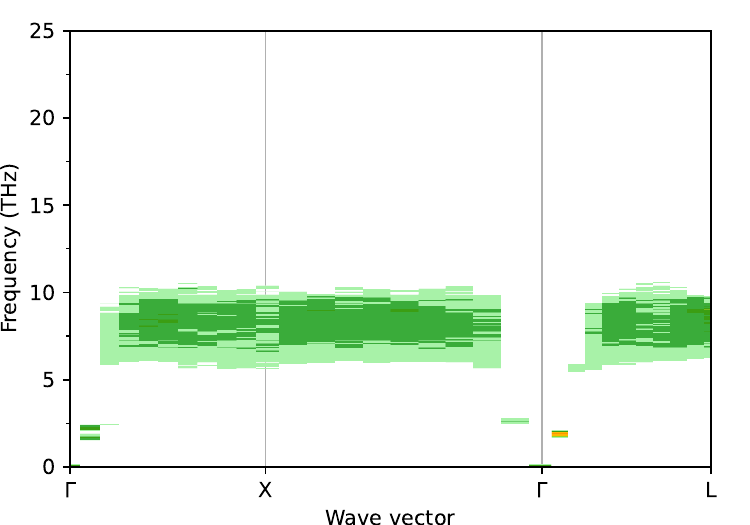}
    \caption{Ti}
    \label{fig:S2c}
  \end{subfigure}
  \hfill
  \begin{subfigure}[b]{0.19\textwidth}
    \centering
    \includegraphics[width=\textwidth]{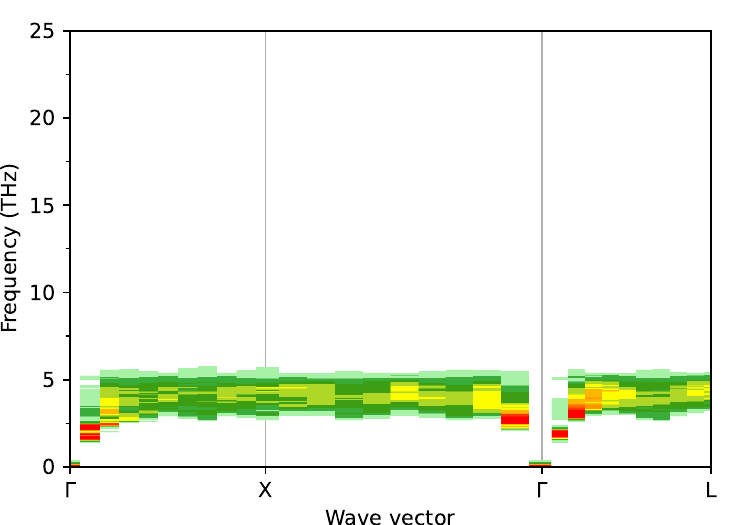}
    \caption{Ta}
    \label{fig:S2d}
  \end{subfigure}
  \hfill
  \begin{subfigure}[b]{0.19\textwidth}
    \centering
    \includegraphics[width=\textwidth]{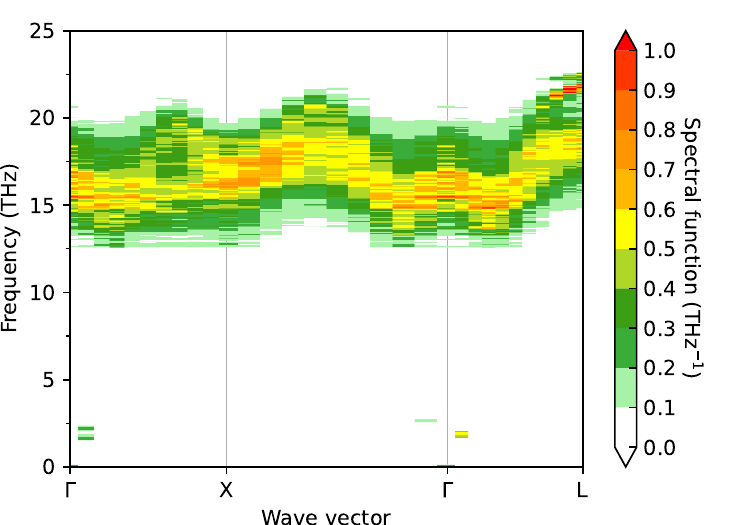}
    \caption{C}
    \label{fig:S2e}
  \end{subfigure}

  \centering
  \begin{subfigure}[b]{0.19\textwidth}
    \centering
    \includegraphics[width=\textwidth]{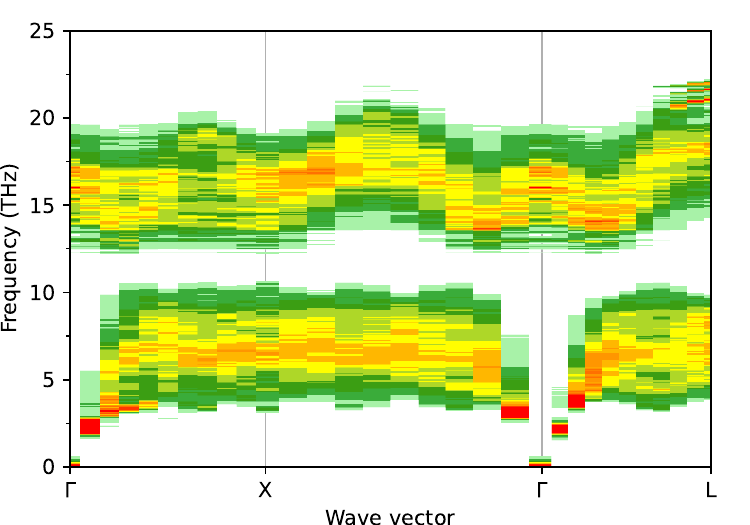}
    \caption{Zr$_{0.33}$Ti$_{0.33}$Nb$_{0.33}$C}
    \label{fig:S2f}
  \end{subfigure}
  \hfill
  \begin{subfigure}[b]{0.19\textwidth}
    \centering
    \includegraphics[width=\textwidth]{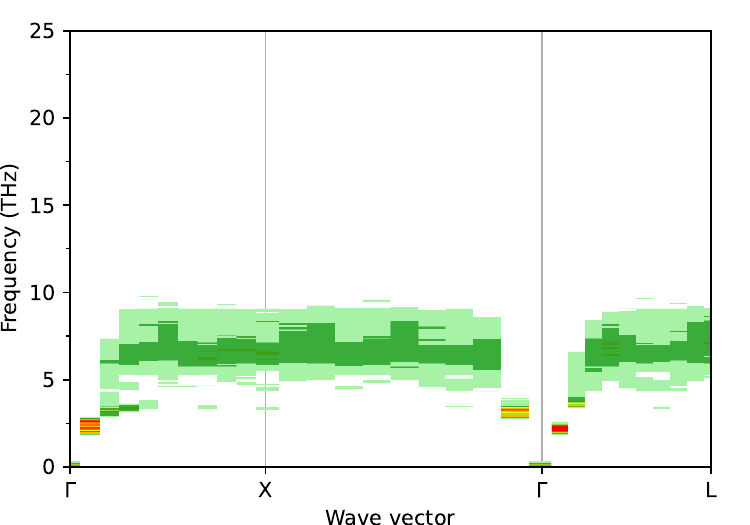}
    \caption{Zr}
    \label{fig:S2g}
  \end{subfigure}
  \hfill
  \begin{subfigure}[b]{0.19\textwidth}
    \centering
    \includegraphics[width=\textwidth]{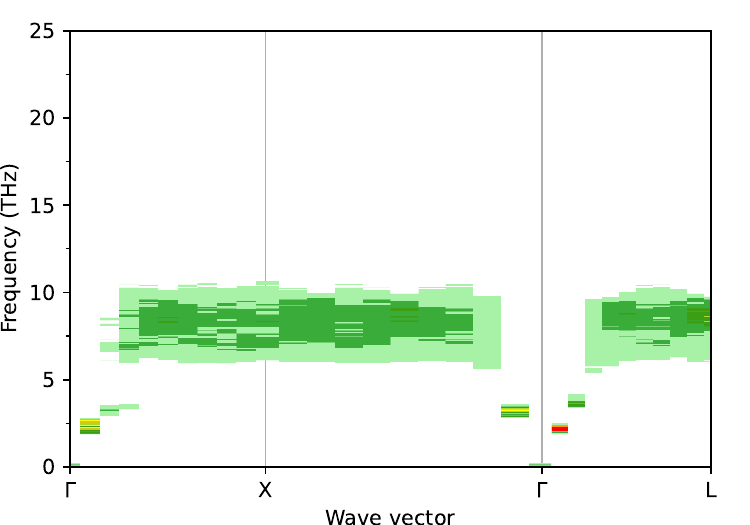}
    \caption{Ti}
    \label{fig:S2h}
  \end{subfigure}
  \hfill
  \begin{subfigure}[b]{0.19\textwidth}
    \centering
    \includegraphics[width=\textwidth]{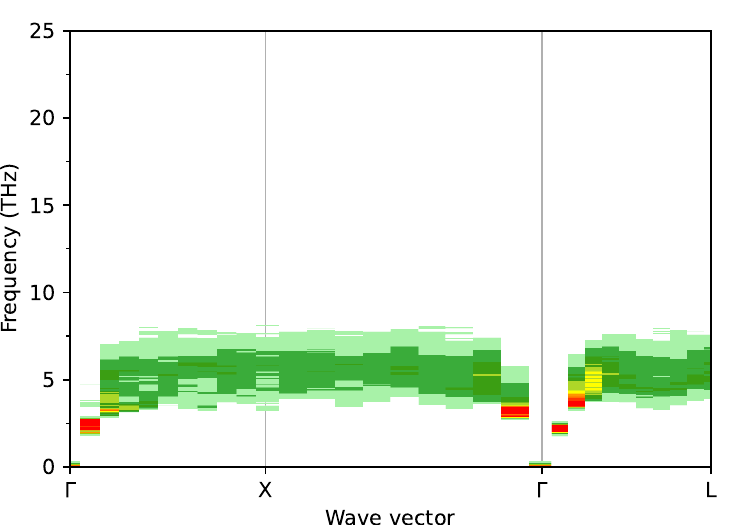}
    \caption{Nb}
    \label{fig:S2i}
  \end{subfigure}
  \hfill
  \begin{subfigure}[b]{0.19\textwidth}
    \centering
    \includegraphics[width=\textwidth]{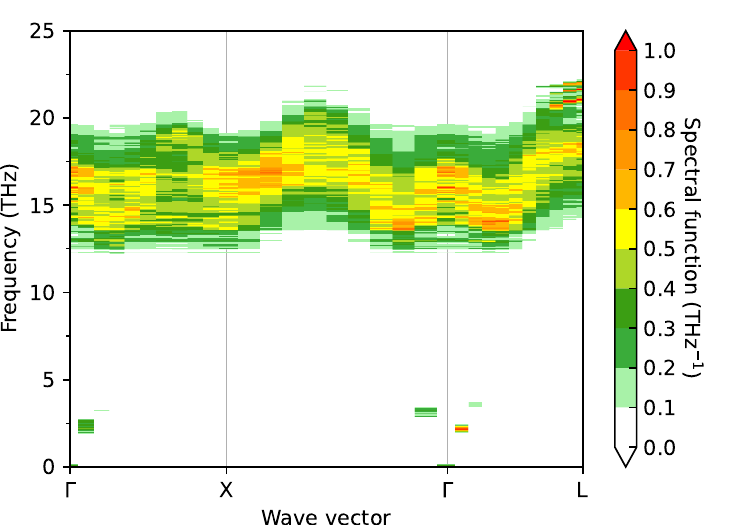}
    \caption{C}
    \label{fig:S2j}
  \end{subfigure}
  \centering
  \begin{subfigure}[b]{0.19\textwidth}
    \centering
    \includegraphics[width=\textwidth]{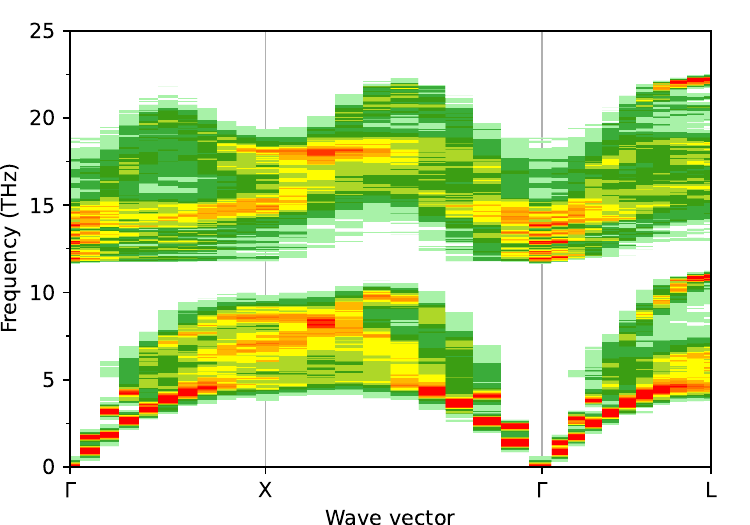}
    \caption{Zr$_{0.33}$Ti$_{0.33}$Hf$_{0.33}$C}
    \label{fig:S2k}
  \end{subfigure}
  \hfill
  \begin{subfigure}[b]{0.19\textwidth}
    \centering
    \includegraphics[width=\textwidth]{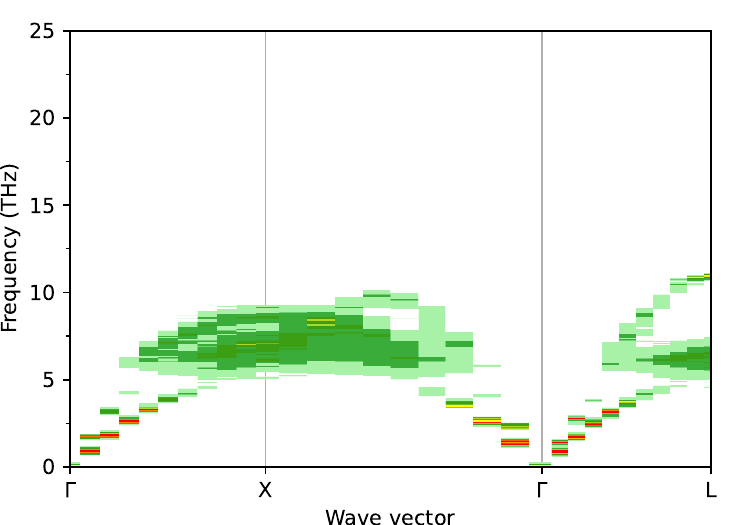}
    \caption{Zr}
    \label{fig:S2l}
  \end{subfigure}
  \hfill
  \begin{subfigure}[b]{0.19\textwidth}
    \centering
    \includegraphics[width=\textwidth]{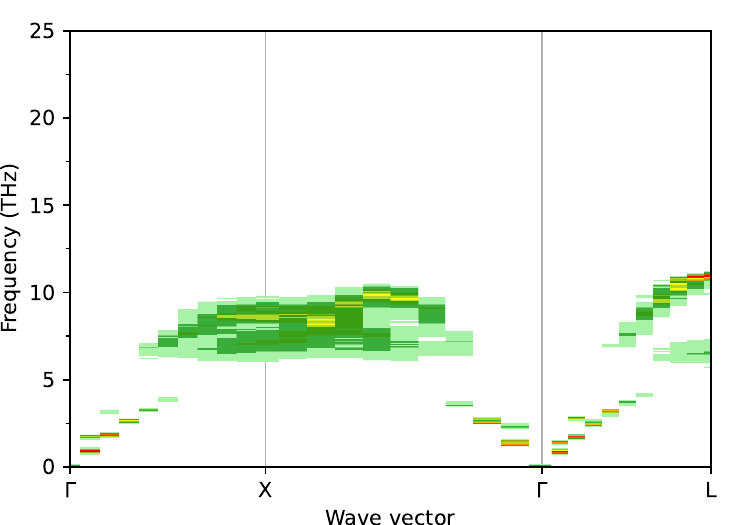}
    \caption{Ti}
    \label{fig:S2m}
  \end{subfigure}
  \hfill
  \begin{subfigure}[b]{0.19\textwidth}
    \centering
    \includegraphics[width=\textwidth]{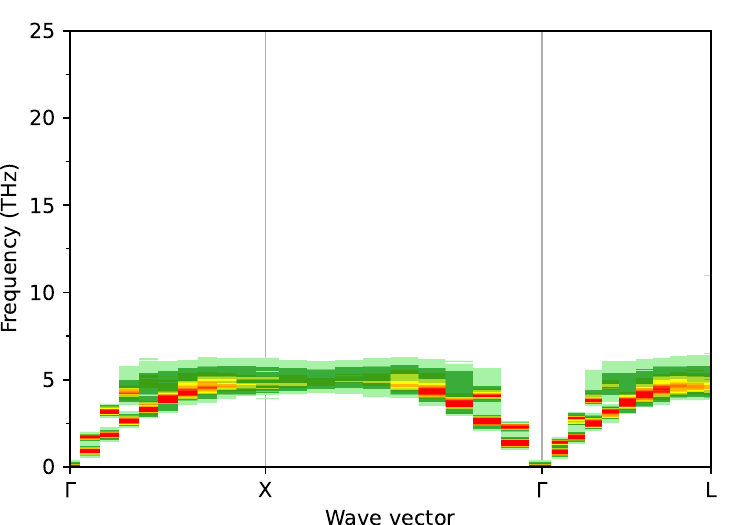}
    \caption{Hf}
    \label{fig:S2n}
  \end{subfigure}
  \hfill
  \begin{subfigure}[b]{0.19\textwidth}
    \centering
    \includegraphics[width=\textwidth]{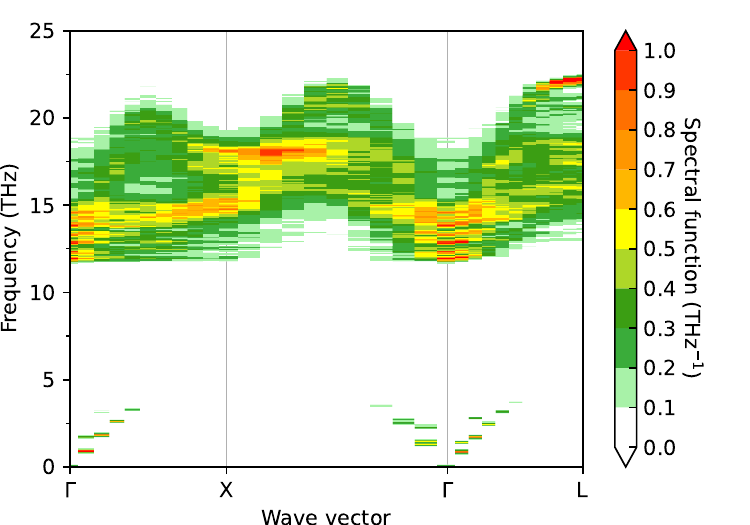}
    \caption{C}
    \label{fig:S2o}
  \end{subfigure}
  
  \centering
  \begin{subfigure}[b]{0.19\textwidth}
    \centering
      \includegraphics[width=\textwidth]{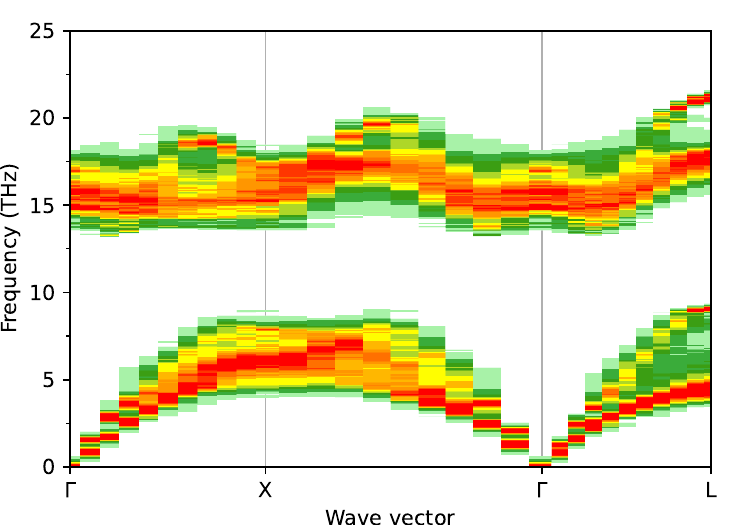}
    \caption{Zr$_{0.33}$Nb$_{0.33}$Hf$_{0.33}$C}
    \label{fig:S2p}
  \end{subfigure}
  \hfill
  \begin{subfigure}[b]{0.19\textwidth}
    \centering
    \includegraphics[width=\textwidth]{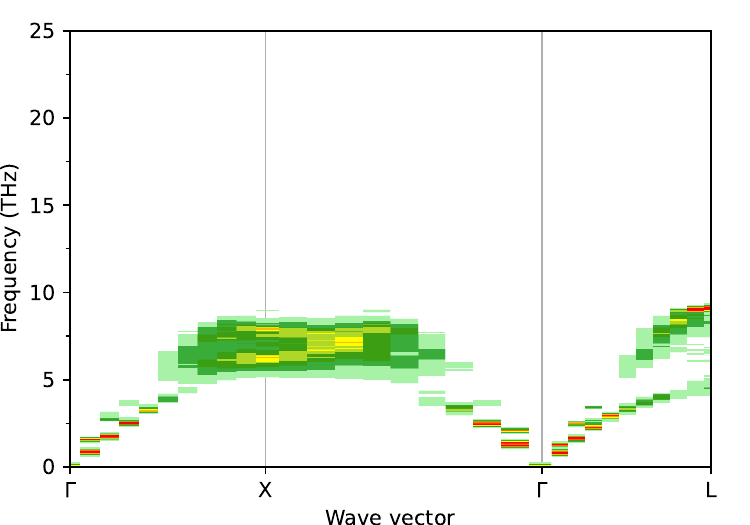}
    \caption{Zr}
    \label{fig:S2q}
  \end{subfigure}
  \hfill
  \begin{subfigure}[b]{0.19\textwidth}
    \centering
    \includegraphics[width=\textwidth]{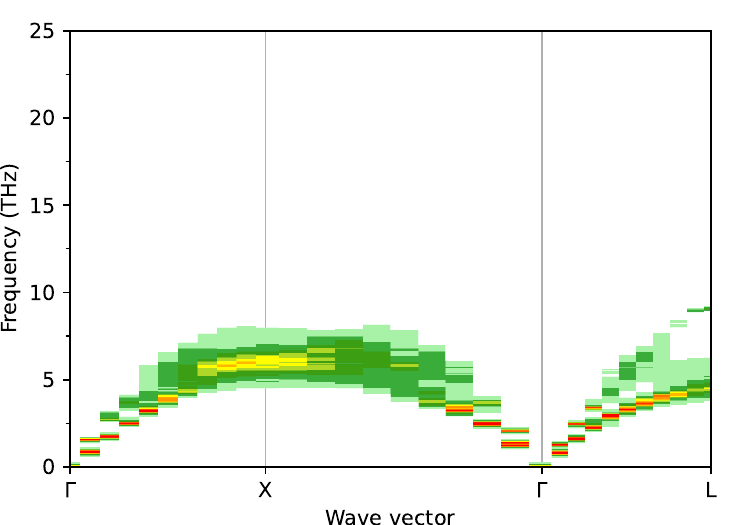}
    \caption{Nb}
    \label{fig:S2r}
  \end{subfigure}
  \hfill
  \begin{subfigure}[b]{0.19\textwidth}
    \centering
    \includegraphics[width=\textwidth]{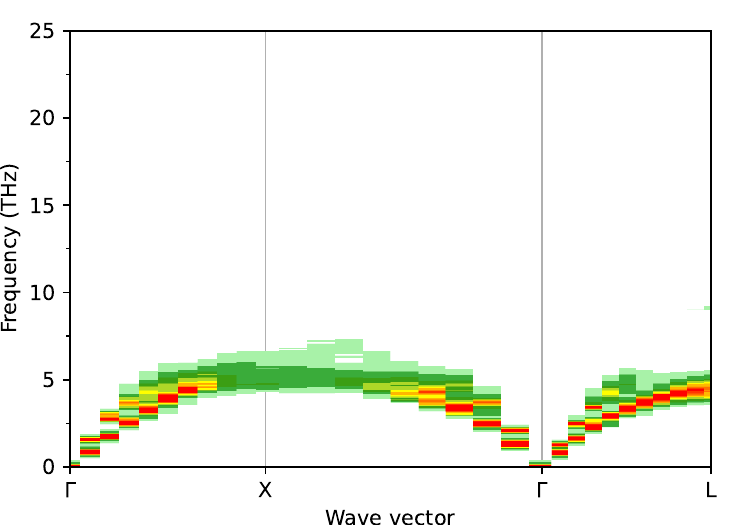}
    \caption{Hf}
    \label{fig:S2s}
  \end{subfigure}
  \hfill
  \begin{subfigure}[b]{0.19\textwidth}
    \centering
    \includegraphics[width=\textwidth]{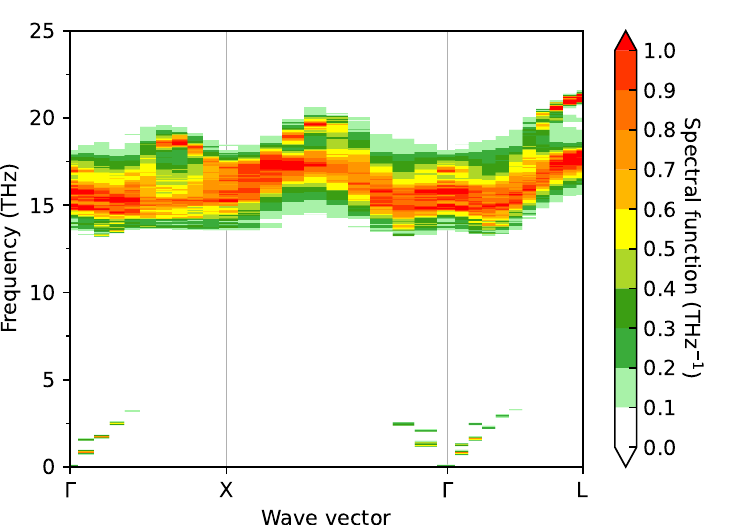}
    \caption{C}
    \label{fig:S2t}
  \end{subfigure}
\caption{\label{fig:S2_overall}Decomposition of the full phonon spectra of the ternary ceramics into the individual elements. The first column is the full phonon spectra and the other columns are the contributions from the individual elements.}
\end{figure*}


\begin{figure*}[ht]
  \centering
  \begin{subfigure}[b]{0.3\textwidth}
    \centering
    \includegraphics[width=\textwidth]{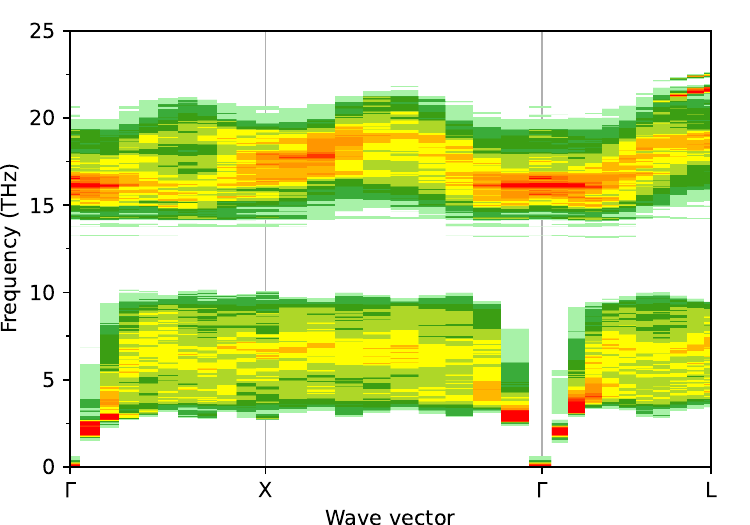}
    \caption{Zr$_{0.25}$Ti$_{0.25}$Ta$_{0.25}$Nb$_{0.25}$C}
    \label{fig:S3a}
  \end{subfigure}
  \hfill
  \begin{subfigure}[b]{0.3\textwidth}
    \centering
    \includegraphics[width=\textwidth]{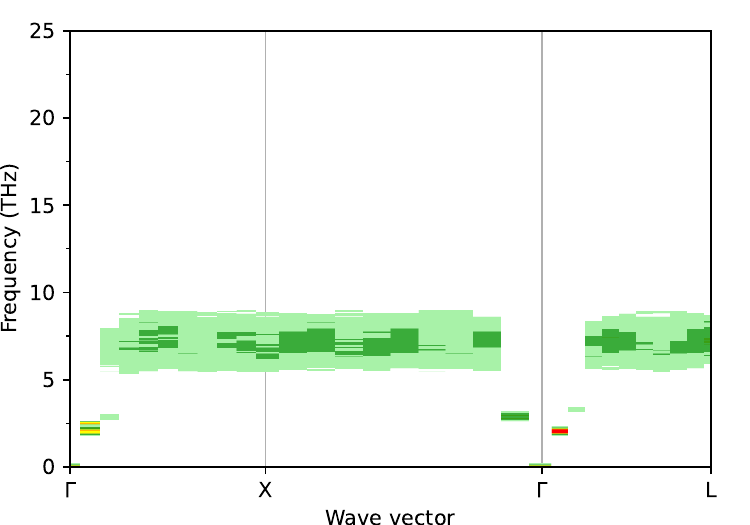}
    \caption{Zr}
    \label{fig:S3b}
  \end{subfigure}
  \hfill
  \begin{subfigure}[b]{0.3\textwidth}
    \centering
    \includegraphics[width=\textwidth]{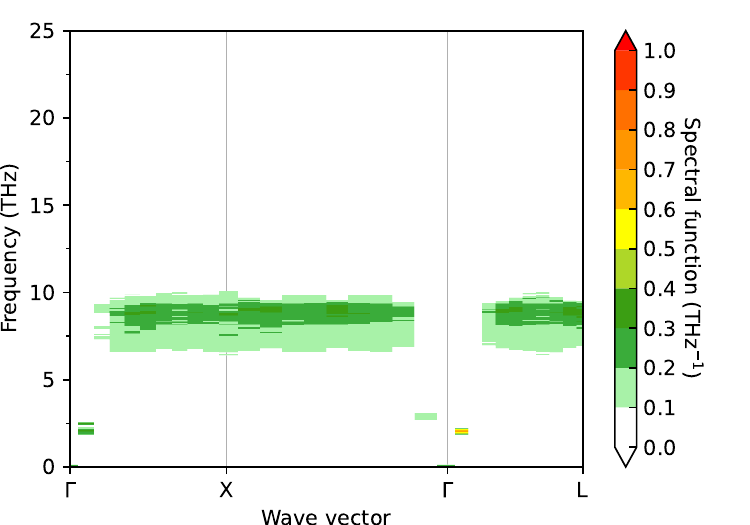}
    \caption{Ti}
    \label{fig:S3c}
  \end{subfigure}
  \hfill
  \begin{subfigure}[b]{0.3\textwidth}
    \centering
    \includegraphics[width=\textwidth]{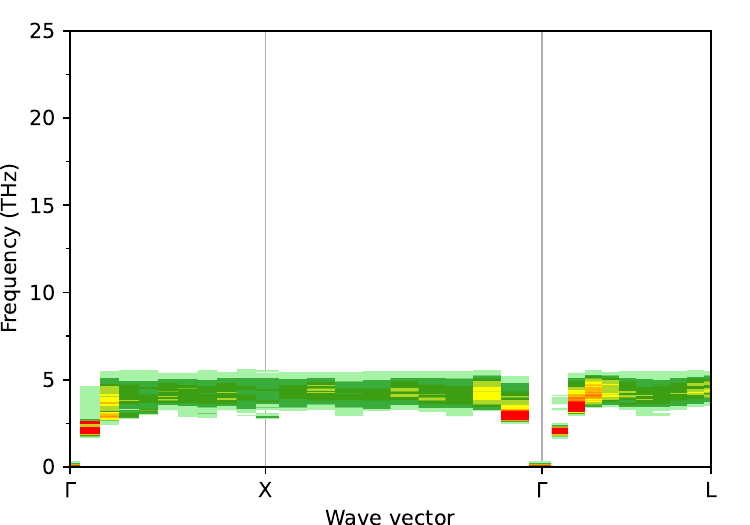}
    \caption{Ta}
    \label{fig:S3d}
  \end{subfigure}
  \hfill
  \begin{subfigure}[b]{0.3\textwidth}
    \centering
    \includegraphics[width=\textwidth]{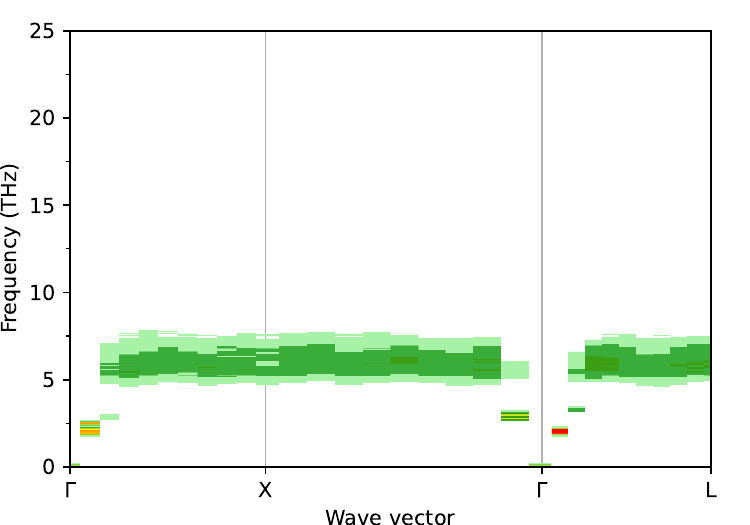}
    \caption{Nb}
    \label{S3e}
  \end{subfigure}
   \hfill
  \begin{subfigure}[b]{0.3\textwidth}
    \centering
    \includegraphics[width=\textwidth]{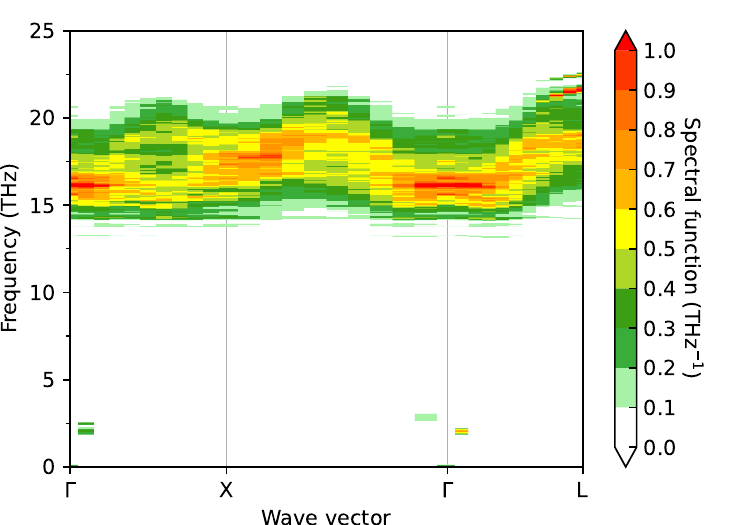}
    \caption{C}
    \label{fig:S3f}
  \end{subfigure}
\caption{\label{fig:S3_overall}Decomposition of the full phonon spectra of the quaternary alloy Zr$_{0.25}$Ti$_{0.25}$Ta$_{0.25}$Nb$_{0.25}$C into individual chemical elements. (a) is the full phonon spectra and (b)-(f) are the contributions from the individual elements.}
\end{figure*}

\begin{figure*}[ht]
  \centering
  \begin{subfigure}[b]{0.3\textwidth}
    \centering
    \includegraphics[width=\textwidth]{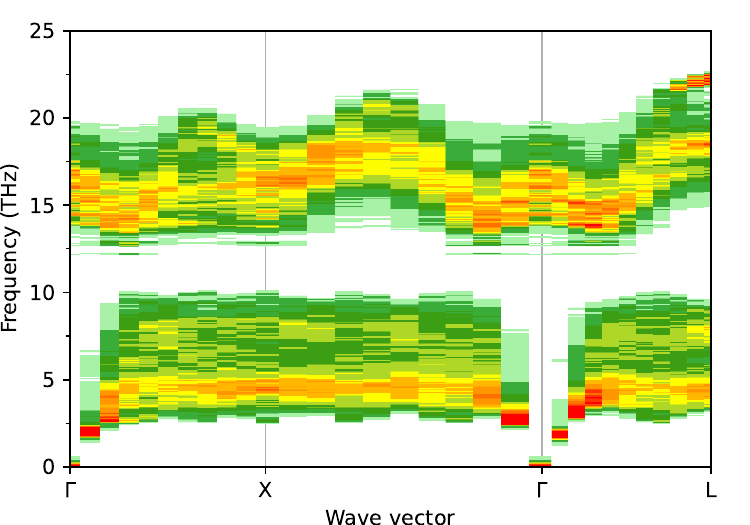}
    \caption{Zr$_{0.25}$Ti$_{0.25}$Ta$_{0.25}$Hf$_{0.25}$C}
    \label{fig:S4a}
  \end{subfigure}
  \hfill
  \begin{subfigure}[b]{0.3\textwidth}
    \centering
    \includegraphics[width=\textwidth]{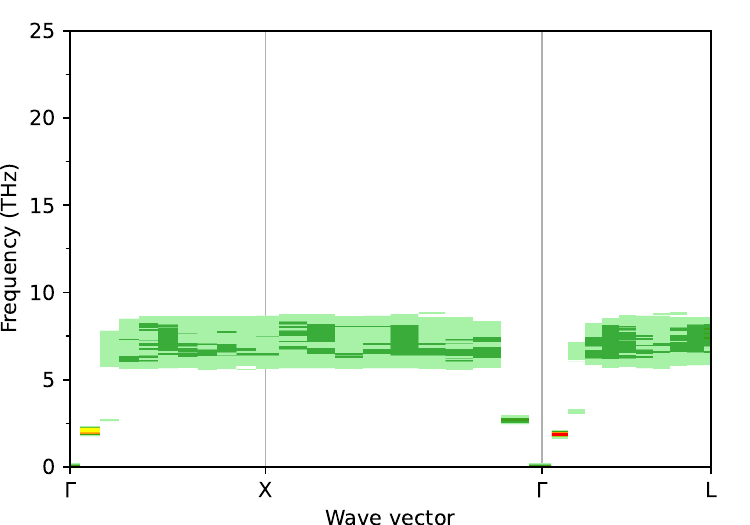}
    \caption{Zr}
    \label{fig:S4b}
  \end{subfigure}
  \hfill
  \begin{subfigure}[b]{0.3\textwidth}
    \centering
    \includegraphics[width=\textwidth]{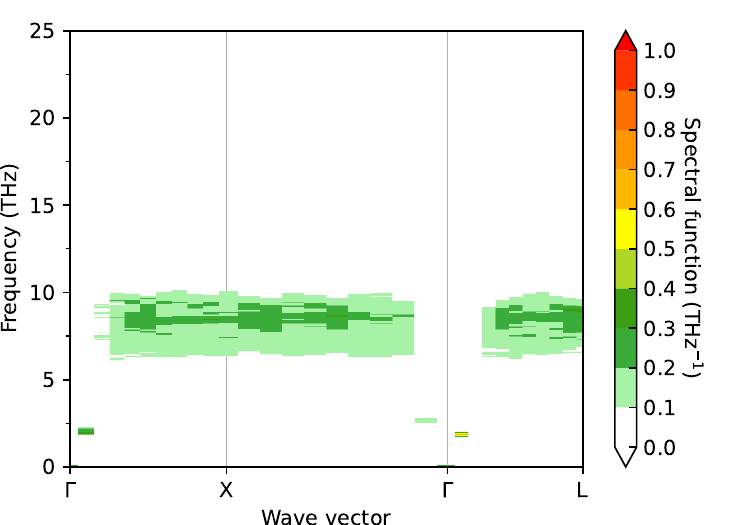}
    \caption{Ti}
    \label{fig:S4c}
  \end{subfigure}
  \hfill
  \begin{subfigure}[b]{0.3\textwidth}
    \centering
    \includegraphics[width=\textwidth]{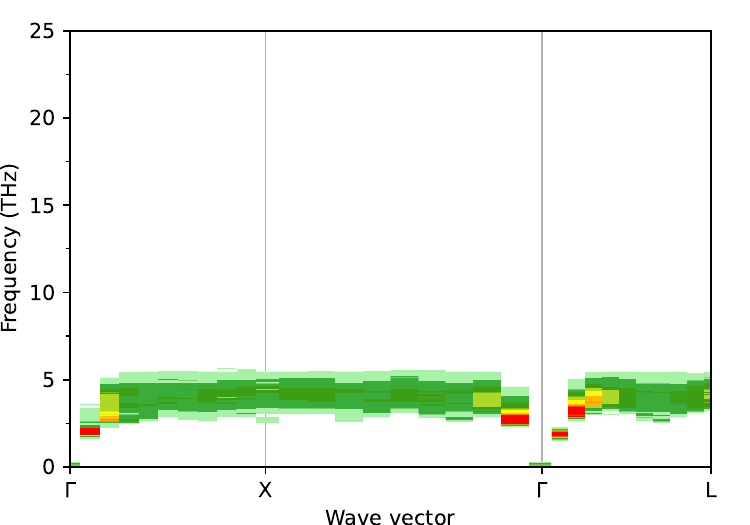}
    \caption{Ta}
    \label{fig:S4d}
  \end{subfigure}
  \hfill
  \begin{subfigure}[b]{0.3\textwidth}
    \centering
    \includegraphics[width=\textwidth]{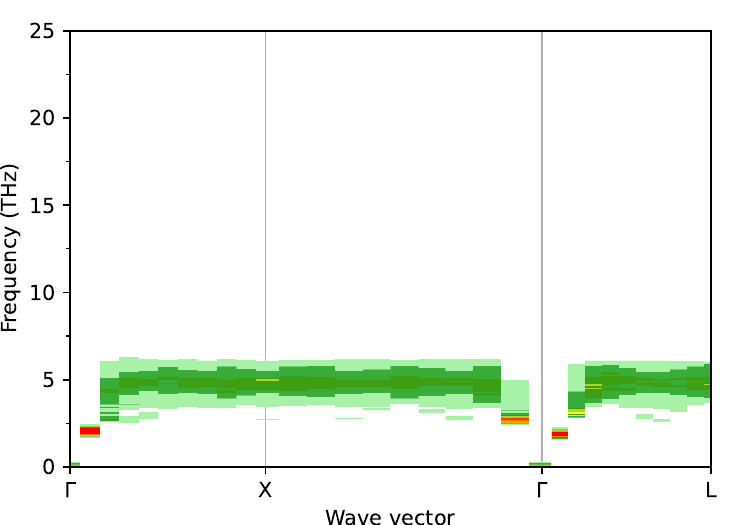}
    \caption{Hf}
    \label{fig:S4e}
  \end{subfigure}
   \hfill
  \begin{subfigure}[b]{0.3\textwidth}
    \centering
    \includegraphics[width=\textwidth]{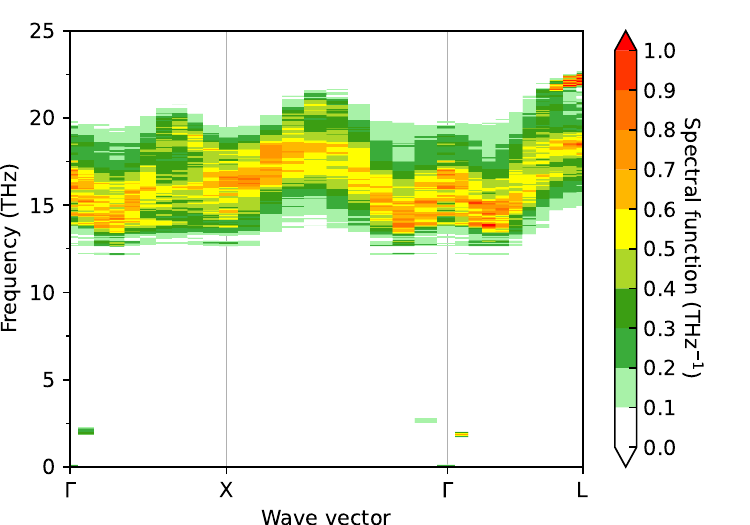}
    \caption{C}
    \label{fig:S4f}
  \end{subfigure}
\caption{\label{fig:S4_overall}Decomposition of the full phonon spectra of the quaternary alloy Zr$_{0.25}$Ti$_{0.25}$Ta$_{0.25}$Hf$_{0.25}$C into individual chemical elements. (a) is the full phonon spectra and (b)-(f) are the contributions from the individual elements.}
\end{figure*}

\begin{figure*}[ht]
  \centering
  \begin{subfigure}[b]{0.3\textwidth}
    \centering
    \includegraphics[width=\textwidth]{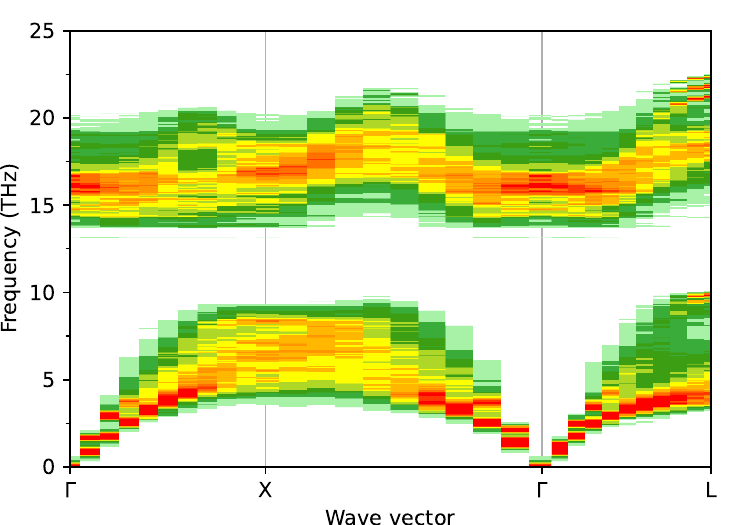}
     \caption{Zr$_{0.2}$Ti$_{0.2}$Ta$_{0.2}$Hf$_{0.2}$Nb$_{0.2}$C}
    \label{fig:S5a}
  \end{subfigure}
  \hfill
  \begin{subfigure}[b]{0.3\textwidth}
    \centering
    \includegraphics[width=\textwidth]{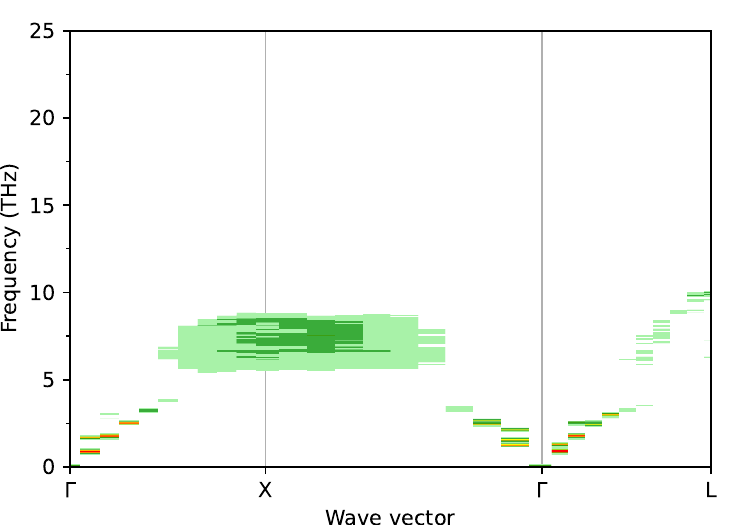}
    \caption{Zr}
    \label{fig:S5b}
  \end{subfigure}
  \hfill
  \begin{subfigure}[b]{0.3\textwidth}
    \centering
    \includegraphics[width=\textwidth]{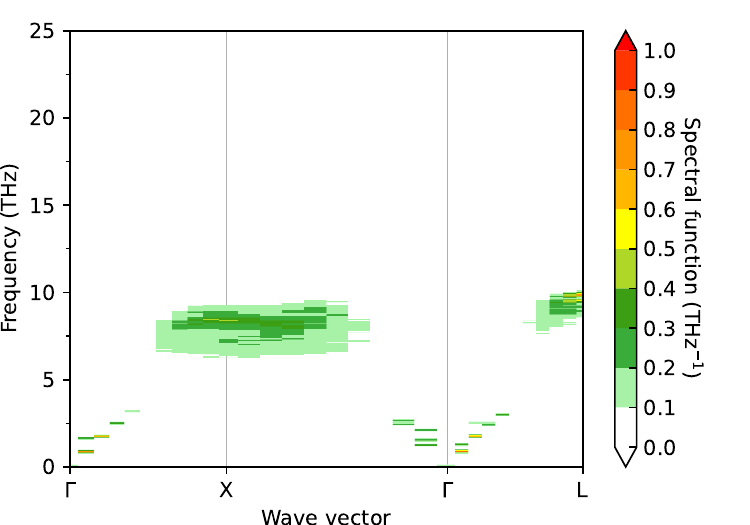}
    \caption{Ti}
    \label{fig:S5c}
  \end{subfigure}
  \hfill
  \begin{subfigure}[b]{0.3\textwidth}
    \centering
    \includegraphics[width=\textwidth]{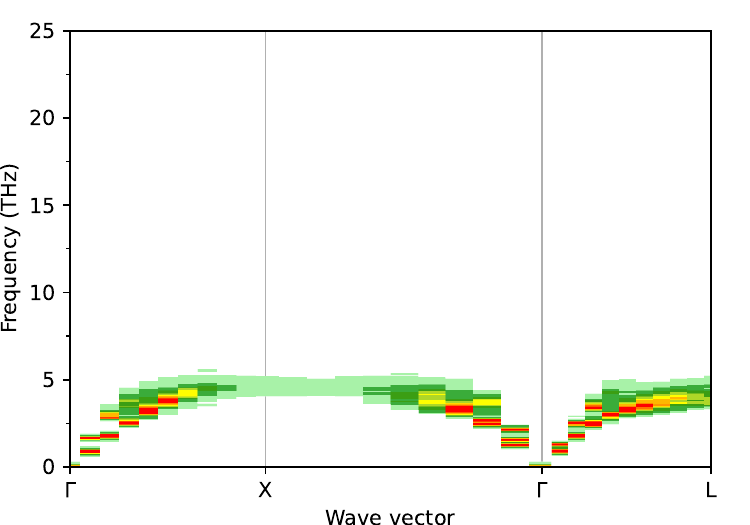}
    \caption{Ta}
    \label{fig:S5e}
  \end{subfigure}
  \hfill
  \begin{subfigure}[b]{0.3\textwidth}
    \centering
    \includegraphics[width=\textwidth]{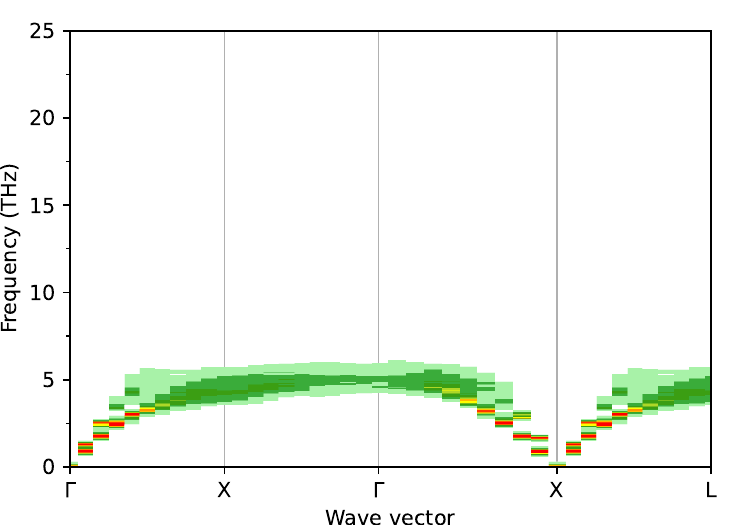}
    \caption{Hf}
    \label{fig:S5f}
  \end{subfigure}
   \hfill
  \begin{subfigure}[b]{0.3\textwidth}
    \centering
    \includegraphics[width=\textwidth]{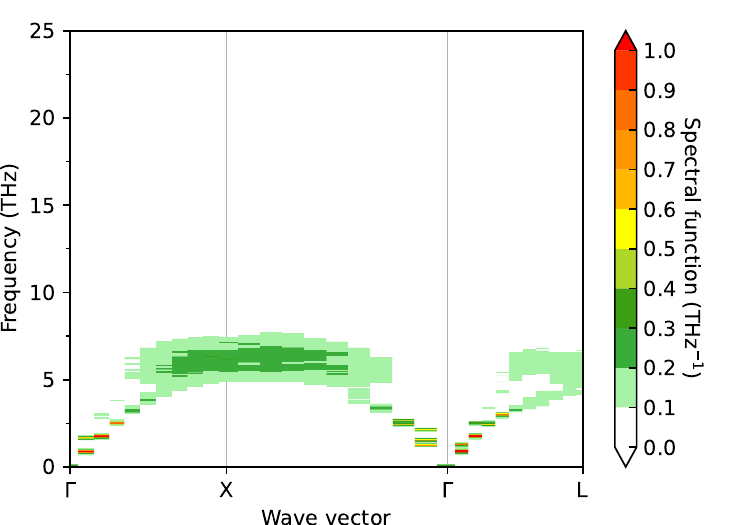}
    \caption{Nb}
    \label{fig:S5g}
  \end{subfigure}
   \hfill
  \begin{subfigure}[b]{0.3\textwidth}
    \centering
    \includegraphics[width=\textwidth]{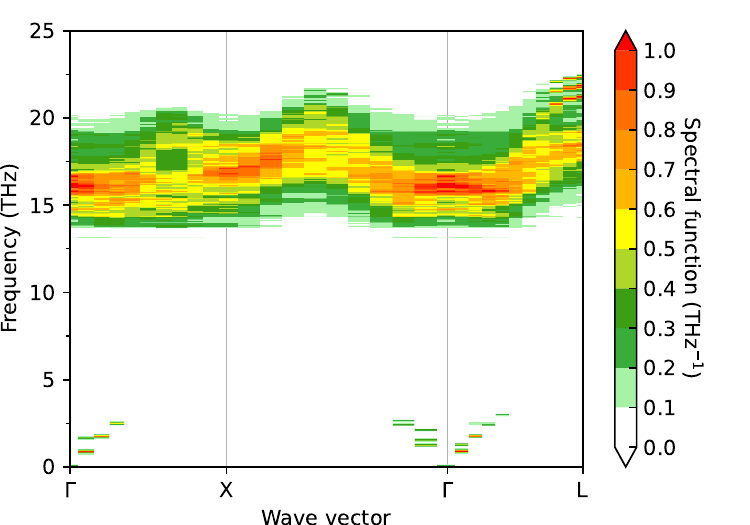}
    \caption{C}
    \label{fig:S5h}
  \end{subfigure}
\caption{\label{fig:S5_overall}Decomposition of the full phonon spectra of the quinary ceramics into the individual chemical elements. (a) is the full phonon spectra and (b)-(g) are the contributions from the individual elements.}
\end{figure*}
\clearpage

\end{document}